\documentclass[aip,pop,preprint]{revtex4-2}
\usepackage{amsmath}
\usepackage{amssymb}
\usepackage{graphicx}
\usepackage{hyperref}
\graphicspath{{figures/}} 
\usepackage[usenames, dvipsnames]{color}

\usepackage[latin1]{inputenc}
\usepackage{tikz}
\usetikzlibrary{trees}
\usetikzlibrary{decorations.pathmorphing}
\usetikzlibrary{decorations.markings}
\usepackage[compat=1.1.0]{tikz-feynman}
\usepackage{soul}

\usepackage{todonotes}
\newcounter{todocounter}

\newcommand{\Ma}{M_{\mathrm{A}}}
\newcommand{\Dk}{D_{\mathrm{Krook}}}
\newcommand{\Pm}{Pm}

\newcommand{\vecu}{\mathbf{u}}
\newcommand{\vecb}{\mathbf{B}}

\newcommand{\veca}{\mathbf{A}}
\newcommand{\vecc}{\mathbf{C}}

\newcommand{\real}{\operatorname{\mathbb{R}e}}

\begin{document}
\author{B.~Tripathi$^1$}
\email{btripathi@wisc.edu}
\author{A.E.~Fraser$^2$}
\author{P.W.~Terry$^1$}
\author{E.G.~Zweibel$^1$}
\author{M.J.~Pueschel$^{3,4}$}
\author{E.H.~Anders$^{5}$}
\affiliation{
$^1$University of Wisconsin-Madison, Madison, Wisconsin 53706, U.S.A.\\
$^2$University of Colorado, Boulder, Colorado 80309, U.S.A.\\
$^3$Dutch Institute for Fundamental Energy Research, 5612 AJ Eindhoven, The Netherlands\\
$^4$Eindhoven University of Technology, 5600 MB Eindhoven, The Netherlands\\
$^5$Center for Interdisciplinary Exploration and Research in Astrophysics, Northwestern University, Evanston, Illinois 60201, U.S.A.
}

\title{Nonlinear mode coupling and energetics of driven magnetized shear-flow turbulence}

\today

\begin{abstract}

To comprehensively understand saturation of two-dimensional ($2$D) magnetized Kelvin-Helmholtz-instability-driven turbulence, energy transfer analysis is extended from the traditional interaction between scales to include eigenmode interactions, by using the nonlinear couplings of linear eigenmodes of the ideal instability. While both kinetic and magnetic energies cascade to small scales, a significant fraction of turbulent energy deposited by unstable modes in the fluctuation spectrum is shown to be re-routed to the conjugate-stable modes at the instability scale. They remove energy from the forward cascade at its inception.  The remaining cascading energy flux is shown to attenuate exponentially at a small scale, dictated by the large-scale stable modes.  Guided by a widely used instability-saturation assumption, a general quasilinear model of instability is tested by retaining all nonlinear interactions except those that couple to the large-scale stable modes. These complex interactions are analytically removed from the magnetohydrodynamic equations using a novel technique. Observations are: an explosive large-scale vortex separation instead of the well-known merger of $2$D, a dramatic enhancement in turbulence level and spectral energy fluxes, and a reduced small-scale dissipation length-scale.  These show critical role of the stable modes in instability saturation. Possible reduced-order turbulence models are proposed for fusion and astrophysical plasmas, based on eigenmode-expanded energy transfer analyses. 

\end{abstract}
\maketitle

\section{Introduction}
\label{sec:intro}

Instability-driven turbulence, commonly found in nature, has traditionally been regarded as similar to externally stirred turbulence, with instability replacing external stirring and nonlinear behavior remaining comparable.\cite{fuller2019, pessah2006, goodman1994, garaud2018, barker2019}  However, insights from studies of instability-driven fusion microturbulence have shown that the two kinds of turbulence are different in many essential regards.\cite{makwana2011, terry2006, terry2021, whelan2018, makwana2014, hatch2011prl, hatch2011, terry2018, li2021, li2022} When turbulence is excited by external forcing at a certain scale, nonlinear interactions between different scales transfer all the injected energy to other scales through an inertial-range energy cascade. In contrast, when an instability taps the free energy of background gradients to drive turbulence, nonlinear interactions quickly become more complex because the nonlinearity excites, at the same instability-scale, other roots of the dispersion relation. This crucially includes linearly stable eigenmodes.\cite{makwana2011, terry2006, terry2021, whelan2018, makwana2014, hatch2011prl, hatch2011, terry2018, li2021, li2022}  Such instability-scale, or large-scale, stable modes are entirely absent in externally stirred turbulence. 
The stable modes, when present, can be excited to a significant level, such that they return turbulent energy from the instability scale to the background gradients, countering the unstable modes that transfer energy in the opposite direction.
This landscape of sources and sinks, mediating the trade of energy between the background gradients and the instability-scale, can have a consequential impact on how small-scale turbulence deals with the energy supplied to it via nonlinear cascades.  
Despite these critical differences between instability-driven and externally-stirred turbulence, it remains the norm to assume that the stable modes do not impact the energetics in magnetohydrodynamic (MHD)\cite{salvesen2014} and fluid turbulence.\cite{smith2021} A careful analysis to test such an assumption is missing.

Stable modes in instability-driven fusion microturbulence mediate the energy injected by the instability at the largest scales excited.  They set the overall fluctuation level and the rates of transport.  Energy is cascaded to small scales, but the amount relative to the energy injected by the instability and removed by the stable modes is so small that it has a negligible effect on transport and fluctuation levels.  
Analysis of saturation of instability-driven microturbulence by stable modes has enabled predictive reduced calculations of transport levels that agree with observations made in comprehensive numerical simulations and experiment,\cite{terry2018} including absolute levels,\cite{terry2021,li2021} the scalings with key parameters,\cite{li2022} and for situations where transport is suppressed above a linear-instability operational-threshold.\cite{whelan2018, terry2021}
Because small-scale cascades are usually of secondary importance for driving turbulent transport---a large-scale aspect, the effect of stable modes on cascades has not been systematically investigated. 

Recently, stable modes have been examined in hydrodynamic and magnetized shear-flow-driven turbulence. It has been established that stable modes are nonlinearly excited,\cite{fraser2017, fraser2018} that they reach levels comparable to those of the linearly unstable modes in saturation, and that they have significant effects on transport.\cite{fraser2021}  In MHD, a critical consideration is the effect of the magnetic field on stable-mode physics in shear-flow-driven turbulence.  The large-scale unstable flow is very efficient at straining magnetic fields aligned with the flow, thus producing a cascade of magnetic energy to small scale.  If the field is sufficiently strong to act back on the flow, and thus on the large-scale stable mode, the smaller scales of the field may blunt the effect of the stable mode.  This effect, which is parameterized by the Alfv\'{e}nic Mach number and which is active somewhat above the instability operational-threshold (i.e., at small Alfv\'{e}nic Mach number),\cite{mak2017} has motivated careful investigation\cite{tripathi2022a} of parametric dependencies of stable-mode excitation. At low Alfv\'{e}nic Mach number, stable modes are impacted, but not to a significant degree.  Despite consequential excitation of stable modes at all strengths of magnetic field, how the small-scale fluctuations, developed via cascades, interact with large-scale stable modes, and how such stable modes set the small-scale fluctuation levels and Kolmogorov dissipative length scales are unknown.  These questions of feedback of large- and small-scales are particularly important as investigating them directly informs if a low-order model of turbulence is feasible.

Here we undertake a systematic investigation of the effect of stable modes on nonlinear energy transfer in two-dimensional shear-flow-driven MHD turbulence.  The investigation necessarily involves connecting the eigenmodes of the Kelvin-Helmholtz (KH) instability with the distinct nonlinearities of MHD---which include the advection of vorticity, the Lorentz force, the advection of field by the flow, and the advection of flow by the field.  The eigenmodes of the ideal KH instability are comprised of an unstable mode and a conjugate stable mode (together labeled as discrete modes henceforth) at a given wavenumber in the unstable range, in addition to a set of neutrally stable modes that form a continuum in frequency.\cite{case1960, fraser2021, tripathi2022a} 
The eigenmodes of the ideal MHD operator are useful as a (complete) set of basis functions because they capture the evolution of turbulence driven by an unstable flow profile and offer physical intuition, e.g., the countering behavior of unstable and stable modes in momentum transport.\cite{tripathi2022b}  Because the discrete modes exist only at large scales, while the continuum modes extend to small scales, the connection between the MHD nonlinearities and the KH eigenmodes is sensitive to scale.  The flow is, however, chiefly concentrated at large scales, while the magnetic field spans a broad range of scales.  The locality or nonlocality of energy transfer is also examined because it further impacts the connections between nonlinearities and eigenmodes.  Analyses of the above effects provide a detailed picture of the relation of stable eigenmodes to cascade directions and different kinds of nonlinear energy transfer functions.

To quantitatively describe the processes indicated in the previous paragraph, nonlinear energy transfer between fluctuations of non-zero wavenumbers requires detailed study. The interactions, mediating energy transfer between fluctuations at three wavenumbers (${\bf k}, {\bf k}', {\bf k}''$), obey the selection rule ${\bf k} = {\bf k}' + {\bf k}''$.   The nonlinearities of energy evolution equations then dictate the transfer, which we write generically as $T({\bf k}\vert {\bf k}',{\bf k}'')$.  This function represents the energy transferred to $\bf k$ from $\bf k'$ and $\bf k''$, collectively. But it is not trivial to identify how the transfer is partitioned among $\bf k'$ and $\bf k''$. Recently,\cite{verma2019} a decomposition of $T({\bf k}\vert {\bf k}',{\bf k}'')$ has been learnt, leading to two rates, $S({\bf k}\vert {\bf k}')$ and $S({\bf k}\vert {\bf k}'')$; the former (latter) $S$-transfer-function uniquely represents signed energy transfer rate to ${\bf k}$ from ${\bf k}'$ (${\bf k}''$) via the intermediary ${\bf k}''$ (${\bf k}'$).  We apply this transfer analysis to MHD shear-flow turbulence, and importantly, render the nonlinear energy transfer in an eigenmode-resolved form to learn about the impact of individual eigenmodes, in particular the stable modes, in MHD cascade processes and instability saturation.

The novel transfer analysis of this paper identifies dominant energy transfer channels from sources of fluctuation energy (unstable modes) to turbulent sinks (stable modes), both lying at large scales.  It reveals whether the saturation-mediating energy transfer is restricted to a subclass of interactions.  This in turn informs, as in the stellarator context,\cite{hegna2018} the important question of whether predictive reduced-order models of the turbulence can be constructed.

This paper is organized as follows. Section~\ref{sec:sec2} details the methods of energy transfer analysis first in wavenumber space and then in eigenmode space. In Sec.~\ref{sec:sec3}, the details of the system set-up are presented. Section~\ref{sec:sec4} shows the numerical analysis of energetic coupling of scales in Kelvin-Helmholtz turbulence. In Sec.~\ref{sec:sec5}, energetic coupling of eigenmodes is analyzed in detail. Findings are related to the interaction of the linear eigenmodes with the evolving profiles of the mean flow and the magnetic field, as well as the nonlinear eigenmode coupling coefficients and nonlinear transfer of energy between the eigenmodes. Section~\ref{sec:sec6} presents a general quasilinear theory of instability saturation and then tests it with a numerical solver, informed by a set of newly derived MHD equations where the large-scale stable modes are analytically removed. Discussion is offered in Sec.~\ref{sec:sec7}.

\section{Machinery for energy transfer analyses}\label{sec:sec2}

An incompressible magneto-fluid evolves according to the standard MHD equations\cite{biskamp2003} as
\begin{subequations}
\begin{align}
    \label{eq:mhdu}
    &\partial_t \vecu = - \vecu \cdot \mathbf{\nabla} \vecu + \vecb \cdot\mathbf{\nabla}\vecb -\nabla\left( P + |\vecb|^2/2\right)  + \nu \nabla^2 \vecu + \mathbf{f},\\ 
    \label{eq:mhdb}
    &\partial_t  \vecb = -\vecu \cdot \mathbf{\nabla} \vecb + \vecb \cdot \mathbf{\nabla} \vecu  + \eta \nabla^2 \vecb,\\
    \label{eq:divu}
    &\mathbf{\nabla} \cdot \vecu=0,\\
    \label{eq:divb}
    &\mathbf{\nabla} \cdot \vecb=0,
\end{align}
\end{subequations}
where $\vecu$, $\vecb$, $P$, $\nu$, $\eta$, and $\mathbf{f}$ respectively stand for the fluid velocity, magnetic field, fluid pressure, viscosity, Ohmic diffusivity, and externally imposed acceleration to the fluid. The factor $4\pi \rho $, with $\rho$ as the fluid density, has been absorbed in the definition of the magnetic field $\vecb$.

A list of important symbols used in this paper is given in Tab.~\ref{tab:t1} on page \pageref{tab:t1}.
\begin{table}
\begin{tabular}{||c | l ||}  
 \hline
Symbol & Meaning  \\ [0.5ex] 
 \hline\hline 
$T_\vecu(k_x\vert k_x', k_x'')$ & Energy transfer rate to $\vecu(k_x)$ from $k_x'$ and $k_x''$  \\ \hline 
$T_\vecb(k_x\vert k_x', k_x'')$ & Energy transfer rate to $\vecb(k_x)$ from $k_x'$ and $k_x''$  \\ \hline 
$S^{\vecu}_{\vecu}(k_x\vert k_x'')$ & Energy transfer rate to $\vecu(k_x)$ from $\vecu(k_x'')$  \\ \hline 
$S^{\vecb}_{\vecu}(k_x\vert k_x'')$ & Energy transfer rate to $\vecu(k_x)$ from $\vecb(k_x'')$  \\ \hline 
$S^{\vecu}_{\vecb}(k_x\vert k_x'')$ & Energy transfer rate to $\vecb(k_x)$ from $\vecu(k_x'')$  \\ \hline 
$S^{\vecb}_{\vecb}(k_x\vert k_x'')$ & Energy transfer rate to $\vecb(k_x)$ from $\vecb(k_x'')$  \\ \hline 
$\vecu\mathrm{<}$ & Wavenumbers of $\vecu$,  lesser than or equal to $k_0$ \\ \hline 
$\vecu\mathrm{>}$ & Wavenumbers of $\vecu$, greater than $k_0$ \\ \hline 
$\vecb\mathrm{<}$ & Wavenumbers of $\vecb$,  lesser than or equal to $k_0$  \\ \hline 
$\vecb\mathrm{>}$ & Wavenumbers of $\vecb$, greater than $k_0$   \\ \hline 
$\Pi^{\vecu<}_{\vecu>}(k_0)$ & Energy flux from $\vecu\mathrm{<}$ to $\vecu\mathrm{>}$ at $k_0$ \\ \hline 
$\Pi^{\vecu<}_{\vecb>}(k_0)$ & Energy flux from $\vecu\mathrm{<}$ to $\vecb\mathrm{>}$ at $k_0$ \\ \hline 
$\Pi^{\vecb<}_{\vecu>}(k_0)$ & Energy flux from $\vecb\mathrm{<}$ to $\vecu\mathrm{>}$ at $k_0$ \\ \hline 
$\Pi^{\vecb<}_{\vecb>}(k_0)$ & Energy flux from $\vecb\mathrm{<}$ to $\vecb\mathrm{>}$ at $k_0$ \\ \hline 
$\beta_j(k_x)$ & Amplitude of the $j^\mathrm{th}$ eigenmode at $k_x$  \\ \hline 
$\gamma_j(k_x)$ & Growth rate of the $j^\mathrm{th}$ eigenmode at $k_x$  \\ \hline 
$C_{jmn}(k_x, k_x')$ & Non-linear mode-coupling coefficient  \\ \hline 
$Q_{j}(k_x)$ & Energy transfer to the $j^\mathrm{th}$ eigenmode at $k_x$ from MHD fields at $k_x=0,t=0$   \\ \hline 
$R_j(k_x)$ & Energy transfer to the $j^\mathrm{th}$ eigenmode at $k_x$ from time-deviation in MHD fields at $k_x=0$    \\  \hline 
$Q_\vecu(k_x)$ & Linear energy drive at $k_x$ of flow field \\ \hline 
$Q_\vecb(k_x)$ & Linear energy drive at $k_x$ of magnetic field \\ [1ex] 
 \hline
\end{tabular}
\caption {List of symbols used in this paper.} \label{tab:t1}
\end{table}

In  what follows we shall consider a two-dimensional system $(x,z)$ with $z$ as an inhomogeneous direction. Thus, the $x$-averaged background profiles of the flow and magnetic fields can have $z$-dependent structures. We shall consider an unstable mean shear flow $\vecu = U_\mathrm{ref}(z) \hat{\mathbf{e}}_x $ with a flow-aligned uniform magnetic field $\vecb = B_0 \hat{\mathbf{e}}_x$.  In Sec.~\ref{sec:sec2A}, we show the method of studying nonlinear scale-interactions between the velocity and magnetic fields when the system is arbitrarily inhomogeneous, by integrating energies along the $z$-axis. Then, in Sec.~\ref{sec:sec2B}, we probe further by decomposing the fluctuations along the $z$-axis in a complete basis of eigenmodes of the linear operator, corresponding to the initial mean flow and magnetic field.

\subsection{Nonlinear scale-interaction analysis} \label{sec:sec2A}

We may compose evolution equations for both the kinetic and magnetic energies, by first writing an evolution equation for $\vecu(k_x)$ and multiplying it with $\vecu^\ast(k_x)$, and then adding a complex conjugate to the resulting equation (and likewise for the magnetic field) to arrive at
\begin{subequations}
\begin{align}
    \partial_t E_\vecu(k_x) &= Q_\vecu(k_x) + \sum_{\substack{k_x'+k_x''=k_x\\ :k_x'\neq 0\ \mathrm{or\ } k_x''\neq 0}} T_\vecu(k_x\vert k_x', k_x'') + \epsilon_f(k_x), \\
    \partial_t E_\vecb(k_x) &= Q_\vecb(k_x) + \sum_{\substack{k_x'+k_x''=k_x\\ :k_x'\neq 0\ \mathrm{or\ } k_x''\neq 0}}  T_\vecb(k_x\vert k_x', k_x''),
\end{align}
\end{subequations}
with $E_\vecu(k_x) = E_\vecu(-k_x) = \langle |\vecu(k_x)|^2 \rangle_z/2$ and $ E_\vecb(k_x) = E_\vecb(-k_x) = \langle |\vecb(k_x)|^2 \rangle_z /2$ as $z$-integrated energies.  Here,
\begin{equation}
    \langle A \rangle_z = \int_0^{L_z} \frac{dz}{L_z}\, A(z)
\end{equation}
and
\begin{equation}
Q_\vecu(k_x) = T_\vecu(k_x\vert 0, k_x) + T_\vecu(k_x\vert k_x, 0) + \nu  \real \Big\{ \langle\vecu^\ast(k_x)  \cdot \nabla^2    \vecu(k_x)\rangle_z \Big\},
\end{equation} 
with
\begin{equation}\label{eq:Tukx}
T_\vecu(k_x\vert k_x', k_x'') = \real \Bigg\{ \Big\langle \vecu^\ast(k_x) \cdot \Big[ -\vecu(k_x') \cdot \nabla'' \vecu(k_x'') + \vecb(k_x') \cdot \nabla'' \vecb(k_x'') \Big] \Big\rangle_z \Bigg\}.
\end{equation}

The signed transfer function $T_\vecu(k_x\vert k_x', k_x'') $ represents the net energy transfer (positive or negative) to a wavenumber $k_x$ from $k_x'$ and $k_x''$ [similarly for $T_\vecb(k_x\vert k_x', k_x'') $], akin to the net energy transfer in hydrodynamics.\cite{kraichnan1959}

Because the velocity is a real observable in physical space $(x,z)$, Hermiticity imposes a symmetry in the energy transfer function: the energy transfer to the velocity field at $k_x$ from $k_x'$ and $k_x''$ via the triad $(k_x, k_x', k_x'')$ with $k_x=k_x'+k_x''$ is equal to the energy transfer to the velocity field at $-k_x$ from $-k_x'$ and $-k_x''$ via the triad $(-k_x, -k_x', -k_x'')$ with $-k_x=-k_x'-k_x''$, mathematically written as
\begin{equation} \label{eq:conservationT1}
T_\vecu(k_x\vert k_x', k_x'') 
= \real \Bigg\{ \Big\langle u_i \Big[ -u_j'^\ast \nabla_j''^\ast u_i''^\ast + B_j'^\ast \nabla_j''^\ast B_i''^\ast  \Big] \Big\rangle_z \Bigg\}
= T_\vecu(\mathrm{-}k_x\vert \mathrm{-}k_x', \mathrm{-}k_x''). 
\end{equation}

It is straightforward to show that the Fourier transform of the gradient of the effective pressure term $\nabla\left(P+|\vecb|^2/2\right)$ in Eq.~\eqref{eq:mhdu}, when dotted with the $-k_x$ component of the velocity and integrated along the $z$-axis, yields zero in the evolution of the total kinetic energy at any wavenumber $k_x$. This is a consequence of the incompressibility of the flow. 

The linear and nonlinear energy transfer rates related to magnetic energy are
\begin{equation}
Q_\vecb(k_x) = T_\vecb(k_x\vert 0, k_x) + T_\vecb(k_x\vert k_x, 0) + \eta \real \Big\{ \langle \vecb^\ast(k_x)  \cdot \nabla^2    \vecb(k_x) \rangle_z \Big\}
\end{equation} 
and
\begin{equation}\label{eq:Tbkx}
T_\vecb(k_x\vert k_x', k_x'') = \real \Bigg\{ \Big\langle \vecb^\ast(k_x) \cdot \Big[ -\vecu(k_x') \cdot \nabla'' \vecb(k_x'') + \vecb(k_x') \cdot \nabla'' \vecu(k_x'') \Big] \Big\rangle_z \Bigg\}.
\end{equation}
A relation similar to Eq.~\eqref{eq:conservationT1} can be derived for the magnetic energy, yielding $T_\vecb(k_x\vert k_x', k_x'')  = T_\vecb(\mathrm{-}k_x\vert \mathrm{-}k_x', \mathrm{-}k_x'')$.

If any three vector fields $\veca,\vecb,$ and $\vecc$---each representing either a velocity field or a magnetic field---at wavenumbers $k_x, k_x',$ and $k_x''$, respectively, interact to drive the field $\veca(k_x)$ via the nonlinear term $\vecb(k_x') \cdot \nabla'' \vecc(k_x'') + \vecc(k_x'') \cdot \nabla' \vecb(k_x') $, where $\nabla'$ and $\nabla''$ serve as reminders to evaluate the gradients at $k_x'$ and $k_x''$, respectively, then the net energy transfer to $\veca(k_x)$ from $\vecb(k_x')$ and $\vecc(k_x'')$ is given as
\begin{equation} \label{eq:threewaveeg}
T_\veca(k_x\vert k_x', k_x'') = \real \Bigg\{\Big\langle \veca^\ast(k_x) \cdot \Big[\vecb(k_x') \cdot \nabla'' \vecc(k_x'') \Big] +  \veca^\ast(k_x) \cdot \Big[\vecc(k_x'') \cdot \nabla' \vecb(k_x') \Big] \Big\rangle_z \Bigg\}.
\end{equation}
Note that this transfer function does not identify the amount of energy transferred to $\veca(k_x)$ from $\vecc(k_x'')$ with $\vecb(k_x')$ acting purely as an intermediary, or the amount of energy transferred to $\veca(k_x)$ from $\vecb(k_x')$ with $\vecc(k_x'')$ acting purely as an intermediary. In order to identify such a wavenumber-to-wavenumber transfer, a few additional transfer functions are useful to define. They will offer convenience in analyzing the energy exchange between different Fourier modes of the velocity and the magnetic fields. We first present a general expression, a priori as a purely mathematical construct,
\begin{equation} \label{eq:scakxkxpp}
    S^{\vecc}_{\veca}(k_x\vert k_x'') =  \real \Bigg\{\Big\langle \veca^\ast(k_x) \cdot \Big[\vecb(k_x') \cdot \nabla'' \vecc(k_x'') \Big] \Big\rangle_z \Bigg\}.
\end{equation}
The subscript and superscript of $S^{\vecc}_{\veca}(k_x\vert k_x'')$ denote $\veca$ at $k_x$ and $\vecc$ at $k_x''$.  
When the fields $\mathbf{A}, \mathbf{B}, \mathbf{C}$ all are divergenceless,
$S^{\vecc}_{\veca}(k_x\vert k_x'')$ satisfies:\cite{verma2019, debliquy2005, teaca2008, teaca2014, teaca2017, grete2017, verma2021}
\begin{subequations}
\begin{align} \label{eq:Santisymm}
S^{\vecc}_{\veca}(k_x\vert k_x'') &=  -S^{\veca}_{\vecc}(k_x''\vert k_x),\\ \label{eq:STrelation}
T_{\veca}(k_x\vert k_x', k_x'')  &=  S^{\vecc}_{\veca}(k_x\vert k_x'') + S^{\vecb}_{\veca}(k_x\vert k_x'),
\end{align}
\end{subequations}
where
\begin{equation}
    S^{\vecb}_{\veca}(k_x\vert k_x') =  \real \Bigg\{\Big\langle \veca^\ast(k_x) \cdot \Big[\vecc(k_x'') \cdot \nabla' \vecb(k_x') \Big] \Big\rangle_z \Bigg\}.
\end{equation}
The proof of Eq.~\eqref{eq:Santisymm} is given in Appendix, and Eq.~\eqref{eq:STrelation} follows from Eq.~\eqref{eq:threewaveeg}.

Because (a) the sum of $S$-transfer-functions yields the net energy transfer function $T$ [see Eq.~\eqref{eq:STrelation}], and, most emphatically, (b) the $S$-transfer-function has the \textit{anti}-symmetry property that the fields $\mathbf{A}$ and $\mathbf{C}$, along with their wavenumbers, can be swapped to gain an overall negative sign [see Eq.~\eqref{eq:Santisymm}], the decomposition of $T$-function into the two $S$-functions has a physical meaning, as has been invoked previously.\cite{verma2019, alexakis2005, debliquy2005, teaca2008, teaca2014, teaca2017, grete2017, verma2021, dong2022}  
The transfer $S^{\vecc}_{\veca}(k_x\vert k_x'')$ physically represents the energy transfer rate to the field $\mathbf{A}$ at $k_x$ from the field $\mathbf{C}$ at $k_x''$, with $\mathbf{B}$ at $k_x'$ acting solely as an intermediary.

Below, we define wavenumber-to-wavenumber $S$-transfer-functions in the context of MHD turbulence\cite{verma2019, alexakis2005, debliquy2005, teaca2008, teaca2014, teaca2017, grete2017, verma2021, dong2022}
\begin{subequations}
\begin{align} \label{eq:suu}
S^{\vecu}_{\vecu}(k_x\vert k_x'') &=  \real \Bigg\{\Big\langle \vecu^\ast(k_x) \cdot \Big[- \vecu(k_x') \cdot \nabla'' \vecu(k_x'') \Big] \Big\rangle_z \Bigg\}, \\ \label{eq:sbu}
S^{\vecb}_{\vecu}(k_x\vert k_x'')  &= \real \Bigg\{ \Big\langle\vecu^\ast(k_x) \cdot \Big[ + \vecb(k_x') \cdot \nabla'' \vecb(k_x'') \Big] \Big\rangle_z \Bigg\}, \\ \label{eq:sub}
S^{\vecu}_{\vecb}(k_x\vert k_x'')  &= \real \Bigg\{ \Big\langle\vecb^\ast(k_x) \cdot \Big[+ \vecb(k_x') \cdot \nabla'' \vecu(k_x'') \Big] \Big\rangle_z \Bigg\}, \\ \label{eq:sbb}
S^{\vecb}_{\vecb}(k_x\vert k_x'')  &= \real \Bigg\{ \Big\langle\vecb^\ast(k_x) \cdot \Big[ -\vecu(k_x') \cdot \nabla'' \vecb(k_x'') \Big] \Big\rangle_z \Bigg\} .
\end{align}
\end{subequations}
We now identify that in a special triad $\left(\vecu(k_x), \vecb(k_x'), \vecb(k_x'') \right) $ the two transfer functions $S^{\vecb}_{\vecu}(k_x\vert k_x'')$ and $S^{\vecu}_{\vecb}(k_x''\vert k_x)$ represent the energy exchange between the velocity at wavenumber $k_x$ and the magnetic field at wavenumber $k_x''$. Since the triad is the same, these two transfer functions are equal in magnitude but of opposite sign, $S^{\vecb}_{\vecu}(k_x\vert k_x'') = -S^{\vecu}_{\vecb}(k_x''\vert k_x)$.

\begin{figure*}
    \centering
    \includegraphics[width=0.9\textwidth]{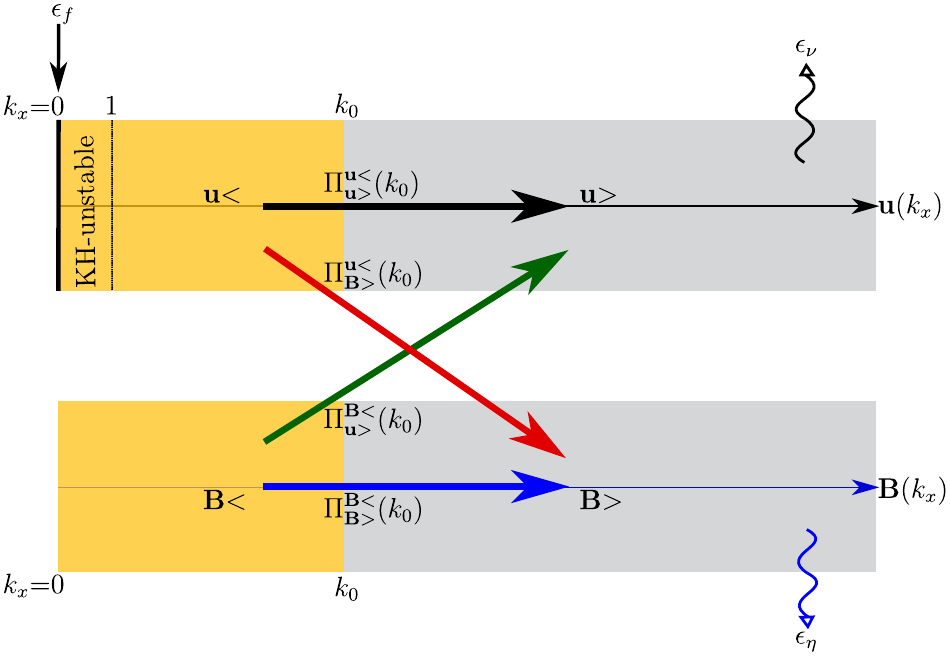}
    \caption{Fourier representation of energy dynamics and cross-scale energy fluxes of MHD turbulence. The top one-dimensional box shows different Fourier modes of the flow $\vecu (k_x)$, whereas the bottom box shows different Fourier modes of the magnetic field $\vecb (k_x)$. The arrows represent energy transfer between different $\vecu (k_x)$ and $\vecb (k_x)$. The arrows need not be in the same directions as shown.  If any of the signed energy fluxes in a physical system is negative, the corresponding arrow direction is reversed, because of the energy conservation, e.g., $-\Pi^{\vecu<}_{\vecu>}(k_0) = \Pi^{\vecu>}_{\vecu<}(k_0)$, where $\vecu\mathrm{<}$ are Fourier modes with wavenumber $k_x \leq k_0$ (yellow colored region). The external forcing $\epsilon_f$ is applied to prevent the relaxation of the mean shear flow, i.e., $\vecu(k_x\mathrm{=}0)$. The wavenumber range $0<|k_x| <1$ represents the large-scale $\mathrm{KH}$ instability of the flow, which drives the turbulence. Small-scale viscous and resistive dissipation are represented by $\epsilon_\nu$ and $\epsilon_\eta$, respectively. }
    \label{fig:f1}
\end{figure*}

Using the above $S$-transfer-functions in MHD turbulence, the cross-scale (signed) energy fluxes passing through a wavenumber $k_0$ are \cite{verma2019, teaca2008, teaca2014, teaca2017, grete2017, verma2021}
\begin{subequations}
    \begin{align} 
        \label{eq:flux1}
        \Pi^{\vecu<}_{\vecu>}(k_0) &= \sum_{|k_x''| \leq k_0} \sum_{|k_x| > k_0} S^{\vecu}_{\vecu}(k_x\vert k_x''),\\  
        \label{eq:flux2}
        \Pi^{\vecu<}_{\vecb>}(k_0) &= \sum_{|k_x''|\leq k_0} \sum_{|k_x| > k_0} S^{\vecu}_{\vecb}(k_x\vert k_x''),\\ 
        \label{eq:flux3}
        \Pi^{\vecb<}_{\vecu>}(k_0) &= \sum_{|k_x''|\leq k_0} \sum_{|k_x| > k_0} S^{\vecb}_{\vecu}(k_x\vert k_x''),\\ 
        \label{eq:flux4}
        \Pi^{\vecb<}_{\vecb>}(k_0) &= \sum_{|k_x''|\leq k_0} \sum_{|k_x| > k_0} S^{\vecb}_{\vecb}(k_x\vert k_x''),
    \end{align}
\end{subequations}
where $\vecu\mathrm{<}$ and $\vecb\mathrm{<}$ represent velocity and magnetic fields at wavenumbers smaller than or equal to $k_0$; similarly, $\vecu\mathrm{>}$ and $\vecb\mathrm{>}$ 
represent respective fields at wavenumbers greater than $k_0$ (Fig.~\ref{fig:f1}). Each flux represents energy flowing in spectral space through $k_0$ from the superscripted index to the subscripted index, e.g., $\Pi^{\vecu<}_{\vecu>}(k_0)$ measures the energy flowing through $k_0$ when the velocity field at wavenumbers less than or equal to $k_0$ transfers energy to the velocity field at wavenumbers greater than $k_0$. Thus, a positive sign of this flux at $k_0$ represents low wavenumbers giving energy to high wavenumbers, i.e., a forward cascade at $k_0$. However, if sign of the flux is found to be negative, it signifies an inverse cascade at $k_0$.

We note that this $z$-integrated formalism alone, while informative with regard to the energy cascade processes in wavenumber space, cannot inform if there are any dominant structures in the fluctuations along the $z$-axis. Analyzing energy transfer between those structures can directly guide reduced-order models of Kelvin-Helmholtz turbulence.  Hence, we next develop a set of novel tools to examine energy transfer between the structures in the fluctuations along the $z$-axis, by decomposing the turbulent fluctuations in a complete basis of the eigenmodes of the linear operator.

\subsection{Nonlinear eigenmode-interaction analysis} \label{sec:sec2B}
To distill the nonlinear interaction between the eigenmodes of the 2$\mathrm{D}$ MHD Kelvin-Helmholtz-instability-driven turbulence, it is advantageous to reduce the number of variables using the streamfunction $\phi$ and flux function $\psi$. Here, $\vecu= \hat{\mathbf{e}}_y \times \nabla \phi$ and $\vecb= \hat{\mathbf{e}}_y \times \nabla \psi$. The governing equations of MHD, presented in Eqs.~\eqref{eq:mhdu}--\eqref{eq:divb}, then can be written as\cite{biskamp2003}
\begin{equation} \label{eq:phispsieqn}
    \begin{aligned}
    \partial_t
    \begin{bmatrix}
    \nabla^2 & 0\\
    0 & 1
    \end{bmatrix}
    \begin{bmatrix}
    \phi\\ 
    \psi
    \end{bmatrix} 
    &= 
    \begin{bmatrix}
    -\{\nabla^2 \phi, \phi \} + \{\nabla^2 \psi, \psi \} + \nu \nabla^4 \phi +  \partial_z f \\ 
    \{\phi, \psi \} + \eta \nabla^2 \psi
    \end{bmatrix},
    \end{aligned}
\end{equation}
where the Poisson bracket is $\{P,Q\} = \partial_x P\cdot \partial_z Q - \partial_z P\cdot \partial_x Q$, e.g., $\{\phi,\psi\} = -\vecu\cdot \nabla \psi$.

We now decompose the streamfunction and magnetic flux function into their mean profiles $[\phi_0, \psi_0]$ at $t=0$, profiles $[\widetilde{\phi}_0, \widetilde{\psi_0}]$ corresponding to deviation of the instantaneous mean from the initial mean, and their fluctuation spectra $[\widetilde{\phi}, \widetilde{\psi}]$ at $k_x \neq 0$. 
Thus, we write a complete decomposition $\phi = \phi_0(k_x\textrm{=}0, t\textrm{=}0) + \widetilde{\phi}_0(k_x\textrm{=}0, t) + \widetilde{\phi}(k_x\neq 0, t) = \phi_0 + \widetilde{\phi}_0 + \widetilde{\phi}$, and likewise for $\psi$. Equation~\eqref{eq:phispsieqn} may now be structurally written, at a fluctuation scale where forcing is not applied, as
\begin{equation} \label{eq:structuraleq}
    \partial_t M \widetilde{X} = L_0 \widetilde{X} + \widetilde{L} \widetilde{X} + L_\mathrm{diss} \widetilde{X} + N (\widetilde{X}, \widetilde{X}) ,
\end{equation}
where $\widetilde{X} = [\widetilde{\phi}, \widetilde{\psi}]^\mathrm{T}$ is the state vector representing the fluctuation spectrum ($k_x \neq 0$); $M$ is a linear operator from Eq.~\eqref{eq:phispsieqn}, given as
\begin{equation} \label{eq:mop}
    M = 
    \begin{bmatrix}
    \nabla^2 & 0\\
    0 & 1
    \end{bmatrix};
\end{equation}
$L_0$ represents the dissipationless linear operator\cite{fraser2021} based on the background profiles prescribed at $t\mathrm{=}0$, and acts on the $\widetilde{X}$ as
\begin{equation}
    L_0 \widetilde{X} = 
    \begin{bmatrix}
    -\{\nabla^2 \phi_0, \widetilde{\phi} \} -\{\nabla^2 \widetilde{\phi}, \phi_0 \} + \{\nabla^2 \psi_0, \widetilde{\psi} \} + \{\nabla^2 \widetilde{\psi}, \psi_0 \} \\ 
    \{\phi_0, \widetilde{\psi} \} + \{\widetilde{\phi}, \psi_0 \}
    \end{bmatrix};
\end{equation}
$\widetilde{L}$ stands for the linear operator formed from the time-fluctuating background profiles ($k_x=0$), and is written as
\begin{equation}
    \widetilde{L}_0 \widetilde{X} = 
    \begin{bmatrix}
    -\{\nabla^2 \widetilde{\phi_0}, \widetilde{\phi} \} -\{\nabla^2 \widetilde{\phi}, \widetilde{\phi_0} \} + \{\nabla^2 \widetilde{\psi_0}, \widetilde{\psi} \} + \{\nabla^2 \widetilde{\psi}, \widetilde{\psi_0} \} \\ 
    \{\widetilde{\phi_0}, \widetilde{\psi} \} + \{\widetilde{\phi}, \widetilde{\psi_0} \}
    \end{bmatrix};
\end{equation}
$L_\mathrm{diss}$ arises from the visco-resistive dissipative terms
\begin{equation}
    L_\mathrm{diss} \widetilde{X} = 
    \begin{bmatrix}
    \nu \nabla^4 \widetilde{\phi} \\ 
    \eta \nabla^2 \widetilde{\psi}
    \end{bmatrix};
\end{equation}
and $N(\widetilde{X},\widetilde{X})$ is a quadratic nonlinear operator (allowing interaction between non-zero Fourier modes)
\begin{equation}
N(\widetilde{X},\widetilde{X})
=
\begin{bmatrix}
    -\{\nabla^2 \widetilde{\phi}, \widetilde{\phi} \} + \{\nabla^2 \widetilde{\psi}, \widetilde{\psi} \} \\ 
    \{\widetilde{\phi}, \widetilde{\psi} \} 
\end{bmatrix}.
\end{equation}

Since the eigenmodes of the linear operator $L_0$ at hand form a complete basis\cite{tripathi2022b}, we may expand at a wavenumber $k_x$ an arbitrary Fourier-transformed state vector $\hat{X}(k_x, z)$ as 
\begin{equation}\label{eq:betaintro}
\hat{X}(k_x,z) = \sum_m \beta_m(k_x) X_m(k_x,z).
\end{equation}
Here, $\hat{X}$ is a Fourier amplitude of $\widetilde{X}$, i.e., $\widetilde{X} = \sum_{k_x \neq 0} \hat{X}(k_x, z) \exp{(i k_x x)}$, and
$X_m(k_x,z)$ is the $m^\mathrm{th}$ eigenmode structure with its complex mode amplitude $\beta_m$, evaluated at $k_x$. We numerically compute well-resolved vertical structures\cite{tripathi2022b} of eigenmodes $X_m(k_x,z)$ using $2048$ Chebyshev polynomials for each variable in an eigenvalue solver, implemented in Dedalus. Note that the eigenmodes $X_m(k_x,z)$ here are non-orthogonal.

We Fourier-transform Eq.~\eqref{eq:structuraleq} and substitute the above eigenmode decomposition for the fluctuations at that wavenumber $k_x$. Then, we multiply the resulting equation with the $j^\mathrm{th}$ left eigenmode $Y_j$ at the same wavenumber. Left eigenmodes are used here because they form a biorthogonal basis with the set of the right eigenmodes $X_m$ in the manner, $\langle Y_j, M X_m\rangle = \delta_{j,m}$. Thus we arrive at

\begin{subequations}
\begin{align}
&\left\langle Y_j, \partial_t M \sum_m \beta_m X_m \right\rangle=  \left\langle Y_j, \left[ L_0  \sum_m \beta_m X_m + \widetilde{L}  \hat{X}+ L_\mathrm{diss} \hat{X} \right] \right\rangle + \sum_{\substack{k_x', k_x''\\:k_x'+k_x''=k_x}} \left\langle Y_j,\hat{N} (\hat{X}', \hat{X}'') \right\rangle,\\
\label{eq:modeeqn1}
&\partial_t  \beta_j = \sum_m \beta_m \left\langle Y_j,  L_0  X_m \right\rangle +\left\langle Y_j, \left[ \widetilde{L}  \hat{X} + L_\mathrm{diss} \hat{X} \right] \right\rangle + \sum_{k_x'} \left\langle Y_j, \hat{N} (\hat{X}', \hat{X}'') \right\rangle,
\end{align}
\end{subequations}
where $\hat{X}'$ and $\hat{X}''$ stand for state vectors at $k_x'$ and $k_x''$ (with a constraint $k_x'+k_x''=k_x$).

As the eigenmodes are obtained for the linear opearator $L_0$, the identity $L_0 X_m = \gamma_m M X_m$ may be used, where $\gamma_m$ is complex eigenvalue corresponding to the $m^\mathrm{th}$ right eigenmode at wavenumber $k_x$. Equation~\eqref{eq:modeeqn1} can thus be simplified, using the orthogonality of the left and the right eigenmodes, to
\begin{equation} \label{eq:betaevoln1}
\partial_t  \beta_j(k_x) = \gamma_j(k_x) \beta_j(k_x) + \left\langle Y_j, \left[\widetilde{L}  \hat{X} + L_\mathrm{diss} \hat{X} \right] \right\rangle+ \sum_{k_x'} \left\langle Y_j, \hat{N} (\hat{X}', \hat{X}'') \right\rangle.
\end{equation}

At this stage, the arbitrary state vector $\hat{X}(k_x,z)$ at wavenumber $k_x$ may be decomposed into a complete set of eigenmodes at that wavenumber. Without loss of generality, in an unstable wavenumber range $|k_x|<1$, we can decompose an arbitrary fluctuation into an unstable $X_1$ eigenmode and a conjugate-stable $X_2$ eigenmode, in addition to the remaining summed fluctuations that encompass the sea of continuum modes $X_\mathrm{c}$.  \cite{tripathi2022b}  This yields
\begin{equation} \label{eq:discretecontinuum}
    \hat{X}(k_x,z) = \left[ \sum_{j=1}^{2} \beta_j(k_x) X_j(k_x,z) \right] + X_\mathrm{c} = X_\mathrm{d} + X_\mathrm{c}.
\end{equation}
The fluctuations corresponding to the unstable and stable modes are the discrete eigenmodes $X_\mathrm{d}$. Substituting Eq.~\eqref{eq:discretecontinuum} in the nonlinearity of Eq.~\eqref{eq:betaevoln1} yields
\begin{equation}
\begin{aligned}[b]
\partial_t \beta_j(k_x) &= \gamma_j(k_x) \beta_j(k_x) + \left\langle Y_j,  \widetilde{L} \hat{X} \right\rangle + \left\langle Y_j,  L_{\mathrm{diss}} \hat{X} \right\rangle \\
&\hspace{2.75cm}+ \sum_{k_x'}\left\langle Y_j,  \left[\hat{N} (X_\mathrm{d}', X_\mathrm{d}'') + \hat{N}  (X_\mathrm{d}', X_\mathrm{c}'') + \hat{N}  (X_\mathrm{c}', X_\mathrm{d}'') + \hat{N}  (X_\mathrm{c}', X_\mathrm{c}'')\right] \right\rangle.
\end{aligned}
\end{equation}
The nonlinear interaction $\hat{N} (X_\mathrm{d}', X_\mathrm{d}'')$ between the two discrete eigenmodes at wavenumbers $k_x'$ and $k_x''$, driving the mode-amplitude $\beta_j(k_x)$, may be further decomposed into individual eigenmode interactions in terms of the unstable and stable modes. This substitution of $X_\mathrm{d}' = \sum_{m=1}^{2} \beta_m' X_m'$ and $X_\mathrm{d}'' = \sum_{n=1}^{2} \beta_n'' X_n''$ leads to
\begin{equation} \label{eq:timematured1}
\begin{aligned}[b]
\partial_t \beta_j(k_x) &= \gamma_j(k_x) \beta_j(k_x) + \left\langle Y_j,  \widetilde{L} \hat{X} \right\rangle  + \left\langle Y_j,  L_{\mathrm{diss}} \hat{X} \right\rangle  \\
&\hspace{0.0cm}+ \sum_{k_x'}\sum_{m=1}^{2} \sum_{n=1}^{2} C_{jmn}(k_x, k_x') \beta_m' \beta_n''  +\sum_{k_x'} \left\langle Y_j,  \left[\hat{N}  (X_\mathrm{d}', X_\mathrm{c}'') + \hat{N}  (X_\mathrm{c}', X_\mathrm{d}'') + \hat{N}  (X_\mathrm{c}', X_\mathrm{c}'')\right] \right\rangle,
\end{aligned}
\end{equation}
where the appearance of the nonlinear mode-coupling coefficient $C_{jmn}(k_x, k_x')= \left\langle Y_j,  \hat{N} (X_\mathrm{m}', X_\mathrm{n}'') \right\rangle$ has been made explicit. The mode-coupling coefficient has five indices---$j,m,n,k_x$ and $k_x'$---denoting the eigenmode $m$ at $k_x'$ interacts with the eigenmode $n$ at $k_x''$ to drive the eigenmode $j$ at $k_x$.  This coefficient measures the strength of a given three-mode overlap.\cite{fraser2017}

It is now straightforward to derive the eigenmode-energy evolution equation. We multiply Eq.~\eqref{eq:timematured1} with $\beta_j^\ast$ and add the complex conjugate of the resulting equation to find
\begin{equation} \label{eq:dtbeta2}
    \partial_t |\beta_j|^2 = Q_j + R_j + D_j + T_{j\mathrm{dd}} +  T_{j\mathrm{dc}} +  T_{j\mathrm{cc}},
\end{equation}
where 
\begin{subequations}
\begin{align} \label{eq:self}
    Q_j &= 2 \gamma_j |\beta_j|^2,\\  \label{eq:cross}
    R_j &= 2 \mathrm{Re\,}\left[ \left\langle Y_j,  \widetilde{L} \hat{X} \right\rangle \beta_j^\ast \right],\\
    D_j &= 2 \mathrm{Re\,}\left[ \left\langle Y_j,  L_{\mathrm{diss}} \hat{X} \right\rangle  \beta_j^\ast \right],\\ \label{eq:Tjmn}
    T_{jmn}(k_x, k_x') &= 2 \mathrm{Re\,}\left[ C_{jmn}(k_x, k_x') \beta_m' \beta_n'' \beta_j^\ast \right],\\ \label{eq:Tjdd}
    T_{j\mathrm{dd}} &= \sum_{k_x'}\sum_{m=1}^{2} \sum_{n=1}^{2} T_{jmn}(k_x, k_x'),\\ \label{eq:Tjdc}
     T_{j\mathrm{dc}} &= 2 \mathrm{Re\,}\left[\sum_{k_x'} \left\langle Y_j,  \left\{ \hat{N} (X_\mathrm{d}', X_\mathrm{c}'') +  \hat{N} (X_\mathrm{c}', X_\mathrm{d}'') \right\} \right\rangle  \beta_j^\ast \right],\\ \label{eq:Tjcc}
     T_{j\mathrm{cc}} &= 2 \mathrm{Re\,}\left[\sum_{k_x'} \left\langle Y_j,  \hat{N} (X_\mathrm{c}', X_\mathrm{c}'') \right\rangle \beta_j^\ast \right].
\end{align}
\end{subequations}
Here, $Q_j$ refers to a source or sink (arising from the linear operator based on the mean profiles at $t=0$); $R_j$ is the response term due to deviation of the instantaneous mean profiles from the initial profiles; $D_j$ is the visco-resistive dissipation term at $k_x$ associated with the $j^\mathrm{th}$ eigenmode; $T_{jmn}(k_x, k_x')$ is the nonlinear energy transfer to mode $j$ at $k_x$ from the interaction between modes $m$ and $n$ at $k_x'$ and $k_x''$, respectively; $T_{j\mathrm{dd}}$ is the energy transfer to mode $j$ at $k_x$ from all the possible nonlinear interactions between the discrete modes, i.e., unstable and stable modes; $T_{j\mathrm{dc}}$ is the energy transfer to mode $j$ at $k_x$ from the all possible nonlinear interactions between the discrete and the continuum eigenmodes; and $T_{j\mathrm{cc}}$ similarly arises from all possible continuum-continuum mode interactions. Because discrete modes of the Kelvin-Helmholtz linear operator exist only at large scales, $|k_x|<1$, the quantity $T_{j\mathrm{cc}}$ captures the feedback of small-scale ($|k_x|>>1$) fluctuations on the $j^\mathrm{th}$ eigenmode, when $j=1$ or $j=2$.

\section{Simulation set-up}\label{sec:sec3}
\subsection{Background profiles and forcing}
We initialize the system with a mean shear flow $\vecu = U_\mathrm{ref}(z) \hat{\mathbf{e}}_x = U_0 \mathrm{tanh}(z/a) \hat{\mathbf{e}}_x$ and a mean magnetic field $\vecb = B_0 \hat{\mathbf{e}}_x$, where $U_0$ and $a$ represent the amplitude and the half-width of the shear flow, respectively. Fluctuation spectra ($k_x \neq 0$) are then excited with low-amplitude, random perturbations such that the energy spectrum in wavenumber space is flat, i.e., no preferential Fourier-mode excitation (for more details, see Section~2 of Ref.~\cite{tripathi2022b}). The perturbed system is then evolved according to Eq.~\eqref{eq:phispsieqn}. With no drive, the system quickly relaxes the mean-flow profile toward a stable configuration. The ensuing turbulence is thus decaying.\cite{fraser2021} To obtain a quasi-stationary state of sustained turbulence, we force the mean flow ($k_x=0$) continuously toward the initial unstable profile with a Krook forcing\cite{smith2021} $\mathbf{f} = f(z) \hat{\mathbf{e}}_x \delta_{k_x,0}$, where
\begin{equation} \label{eq:forcing}
     f(z) = \Dk \left[U_{\mathrm{ref}}(z) - \langle u_x(x,z,t) \rangle_x\right] + F_0,
\end{equation}
where $\Dk$ is the profile relaxation rate \cite{allawala2020, marston2008} that controls the forcing strength, and $\langle u_x(x,z,t) \rangle_x$ is the instantaneous $x$-averaged flow. The time-independent force $F_0$ is imposed only to balance the pure viscous diffusion of the initial shear flow, $\nu \nabla^2 U_{\mathrm{ref}}(z)  + F_0 = 0$, ensuring that, at $t=0$, we realize an initial equilibrium state, about which small-amplitude initial perturbations evolve.\cite{tripathi2022b}

\subsection{Non-dimensionalization}
We non-dimensionalize all variables with length scale $a$ and flow speed $U_0$. Thus, time is measured in units of $a/U_0$, energy (per unit mass) has units of $U_0^2$, and $\Dk$ is specified in terms of $U_0/a$. The mean flow then becomes $U_\mathrm{ref}(z) = \mathrm{tanh}(z)$. The magnetic field strength is quantified by the Alfv\'{e}nic Mach number $\Ma= U_0 /B_0$.  The effects of the viscosity and resistivity are measured by the fluid Reynolds number $Re=aU_0/\nu$ and the magnetic Reynolds number $Rm=aU_0/\eta$, respectively.  The magnetic Prandtl number is defined as $\Pm=Rm/Re$. 

All simulations use a box size of $(L_x, L_z) = (10\pi, 20\pi)$, $Re=500$ (unless stated otherwise), and a high spectral resolution of $2048$ Fourier modes along the $x$-axis and $2048$ Chebyshev polynomials along the $z$-axis to obtain well-converged results.\cite{tripathi2022b}  Additionally, we dealias the quadratic nonlinearities of the system using $3/2$ times the mentioned spectral modes. We employ a pseudospectral numerical solver Dedalus.\cite{burns2020, lecoanet2016} 

Boundary conditions used are, along the $x$-axis, periodic, and, along the $z$-axis, perfectly conducting, no-slip walls, co-moving with the initial flow at the top ($z=L_z/2$) and bottom ($z=-L_z/2$) layers.\cite{tripathi2022b}

\section{Energetic coupling of scales}\label{sec:sec4}

Before we present the saturation properties of the Kelvin-Helmholtz-instability-driven turbulence in terms of the nonlinear interaction between the linear eigenmodes, we analyze the interaction between scales associated with the velocity and magnetic fields. The scale-interaction\cite{verma2019} is usually applied to homogeneous turbulence. Here we modify it for inhomogeneous turbulence.

\subsection{Scale-interaction of velocity and magnetic fields}

\begin{figure*}
    \centering
    \includegraphics[width=0.55\textwidth]{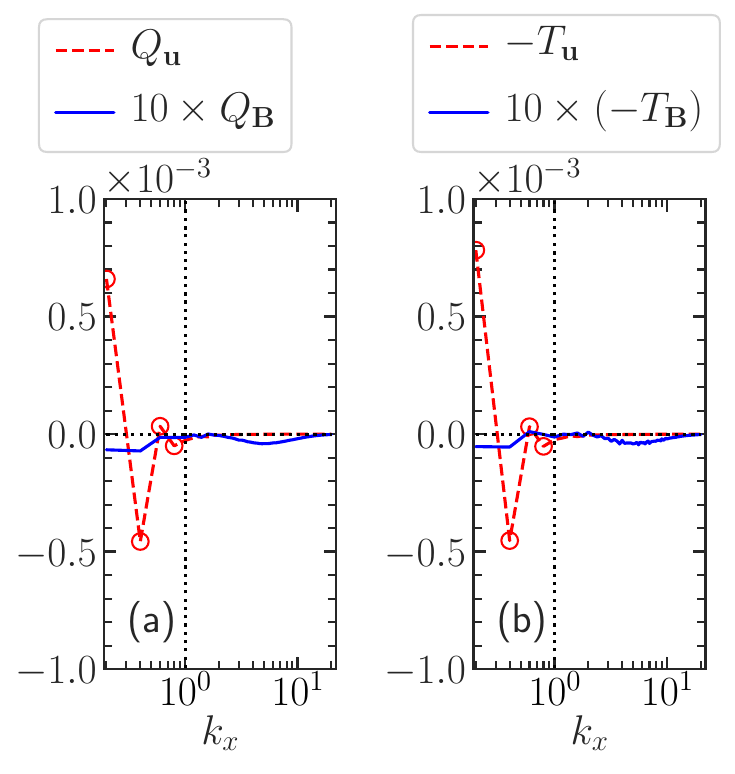}
    \caption{Time-averaged (a) linear and (b) nonlinear energy transfer rates in velocity and magnetic fields. The wavenumbers $|k_x| <1$, to the left of the dotted vertical lines, are Kelvin-Helmholtz-unstable. The time-averaged rates of linear and nonlinear transfers are almost equal and opposite. Note the negative linear energy injection in the flow $Q_\mathbf{u}$ at $k_x=0.4$, despite this being the wavenumber where perturbations linearly grow the fastest. The simulation parameters used are $\Ma=30$, $\Dk=2$, and $\Pm=1$.}
    \label{fig:f2}
\end{figure*}

Because the flow and the magnetic field have mean profiles, it is useful to compare the linear and nonlinear processes in the energy dynamics. The linear energy injection  or removal of energy at each scale is shown in Fig.~\ref{fig:f2}(a), where the presented $Q_\vecu$ and $Q_\vecb$ are time-averaged over a long quasi-stationary state. As expected, the energy is injected only at scales that lie within the Kelvin-Helmholtz-unstable wavenumber range $0 < |k_x| <1$, whereas energy from the fluctuation spectrum is removed via linear process at scales beyond the KH-unstable range. It is also observed, somewhat surprisingly, that the energy is linearly removed from wavenumbers $k_x=0.4$ and $k_x=0.8$ (the second and the fourth Fourier modes). We note that $k_x=0.4$ is the wavenumber that is most unstable linearly. The reason behind this energy removal will later be explained by analyzing the energy transfer at each scale, decomposing the fluctuation into different eigenmodes instead of integrating the fluctuation spectrum along the $z$-axis, as we have done here. 

\begin{figure*}
    \centering
    \includegraphics[width=1\textwidth]{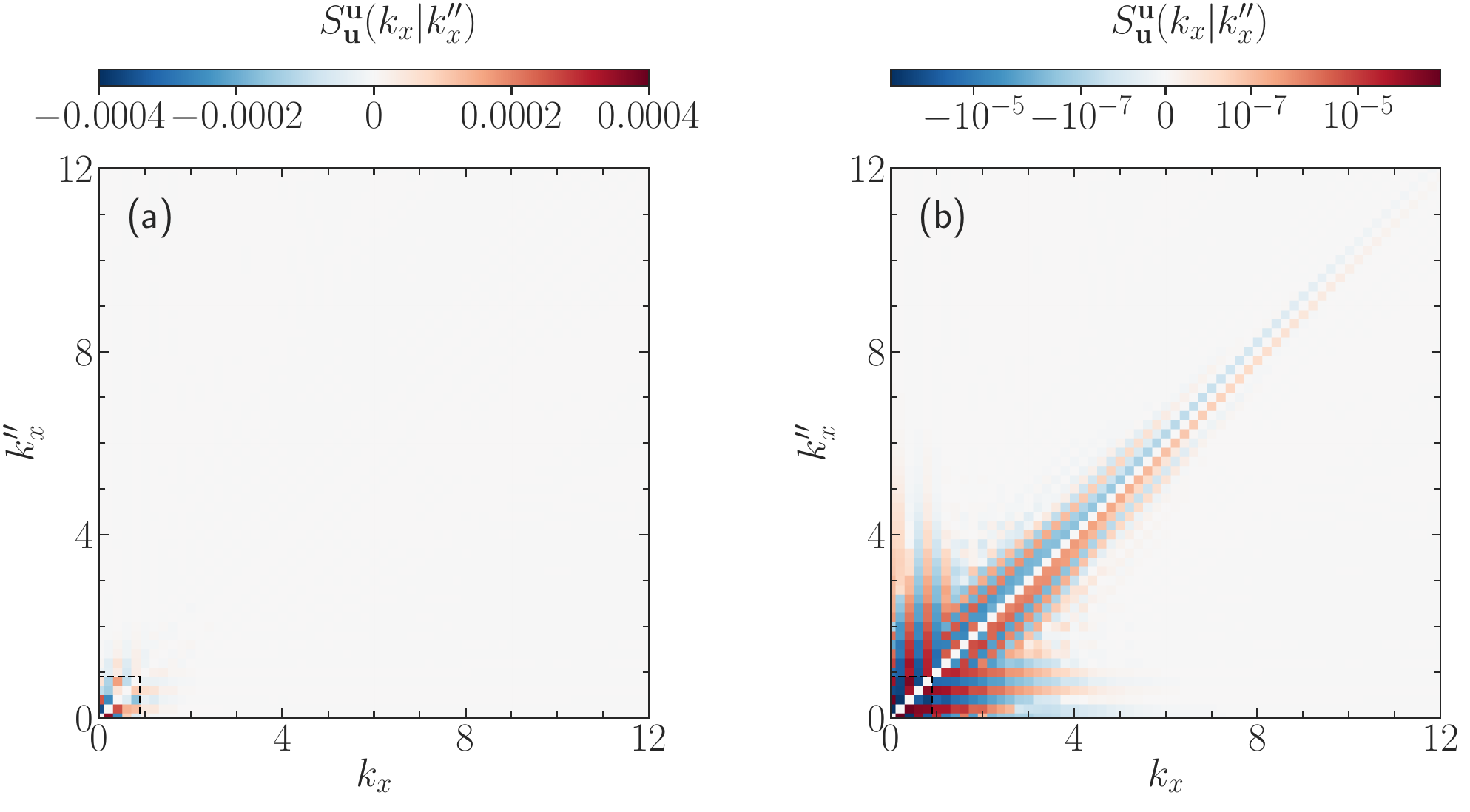}
    \caption{Time-averaged wavenumber-to-wavenumber energy transfer rates among the velocity fields. (a) The transfer is almost entirely localized within the Kelvin-Helmholtz instability range, shown with a black dashed box near the lower leftmost end of each subplot. (b) Logarithmic spectrum of the energy transfer reveals nonlocal triads, but local energy transfer, enabled by the box-sized Kelvin-Helmholtz eddy, i.e., the first non-zero Fourier mode number.  The simulation parameters used are $\Ma=120$, $\Dk=2$, and $\Pm=1$.}
    \label{fig:f3}
\end{figure*}

\begin{figure*}
    \centering
    \includegraphics[width=0.75\textwidth]{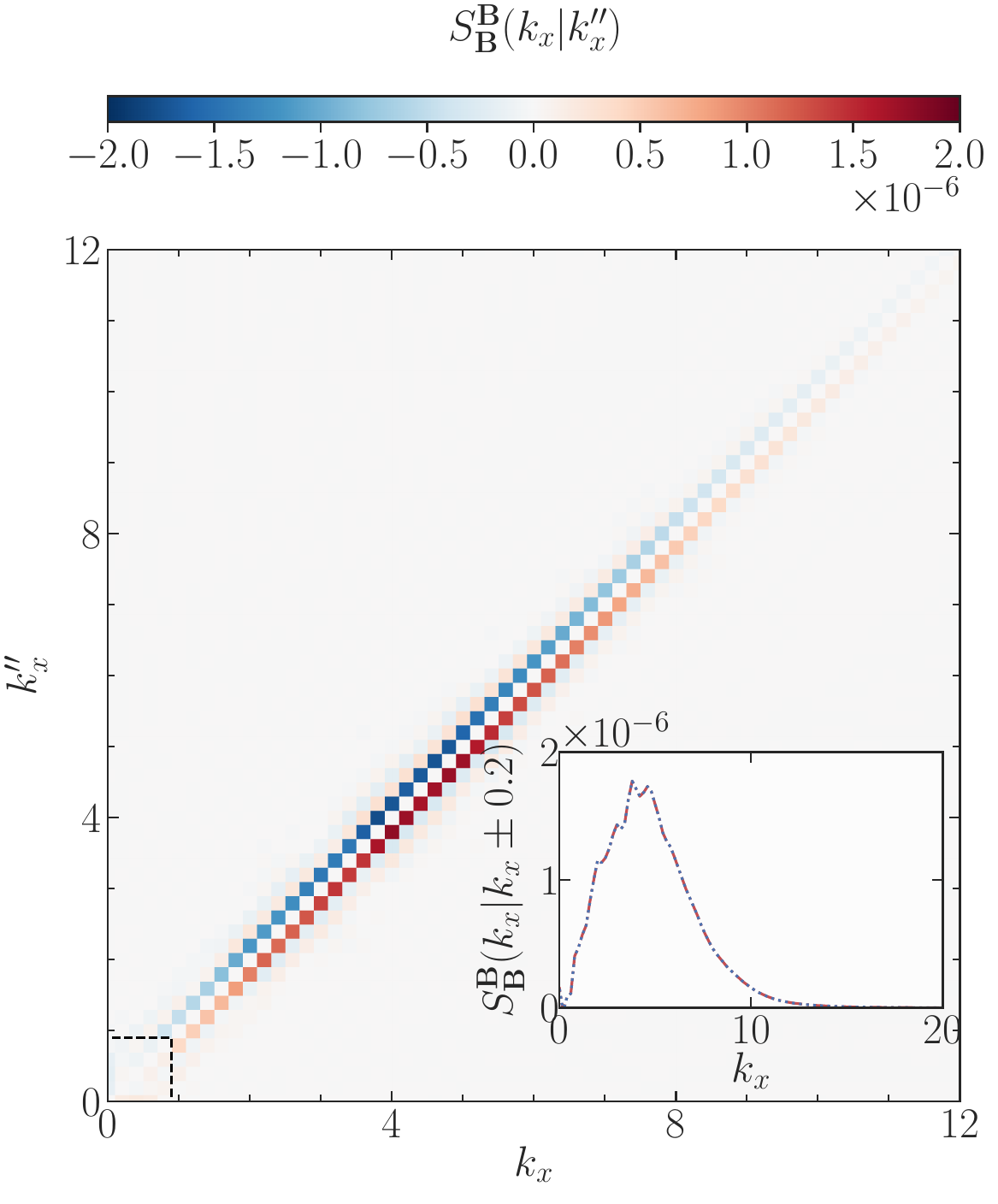}
    \caption{Time-averaged wavenumber-to-wavenumber energy transfer rates among the magnetic fields. The transfer is dominant outside the Kelvin-Helmholtz instability range, shown with a black dashed box near the lower leftmost end of the plot. The transfer $\vecb$-to-$\vecb$ involves nonlocal triads but energy is locally transferred---from a high wavenumber to another high wavenumber. Shown in the inset are one-dimensional spectra of the transfer function, following the two diagonals, $k_x''-k_x = \pm 2\pi/L_x$. The two curves, red and black, are identical, although one is positive and another negative in sign, which is a consequence of energy conservation in a triad. The simulation parameters used are $\Ma=120$, $\Dk=2$, and $\Pm=1$.}
    \label{fig:f4}
\end{figure*}

\begin{figure*}
    \centering
    \includegraphics[width=1\textwidth]{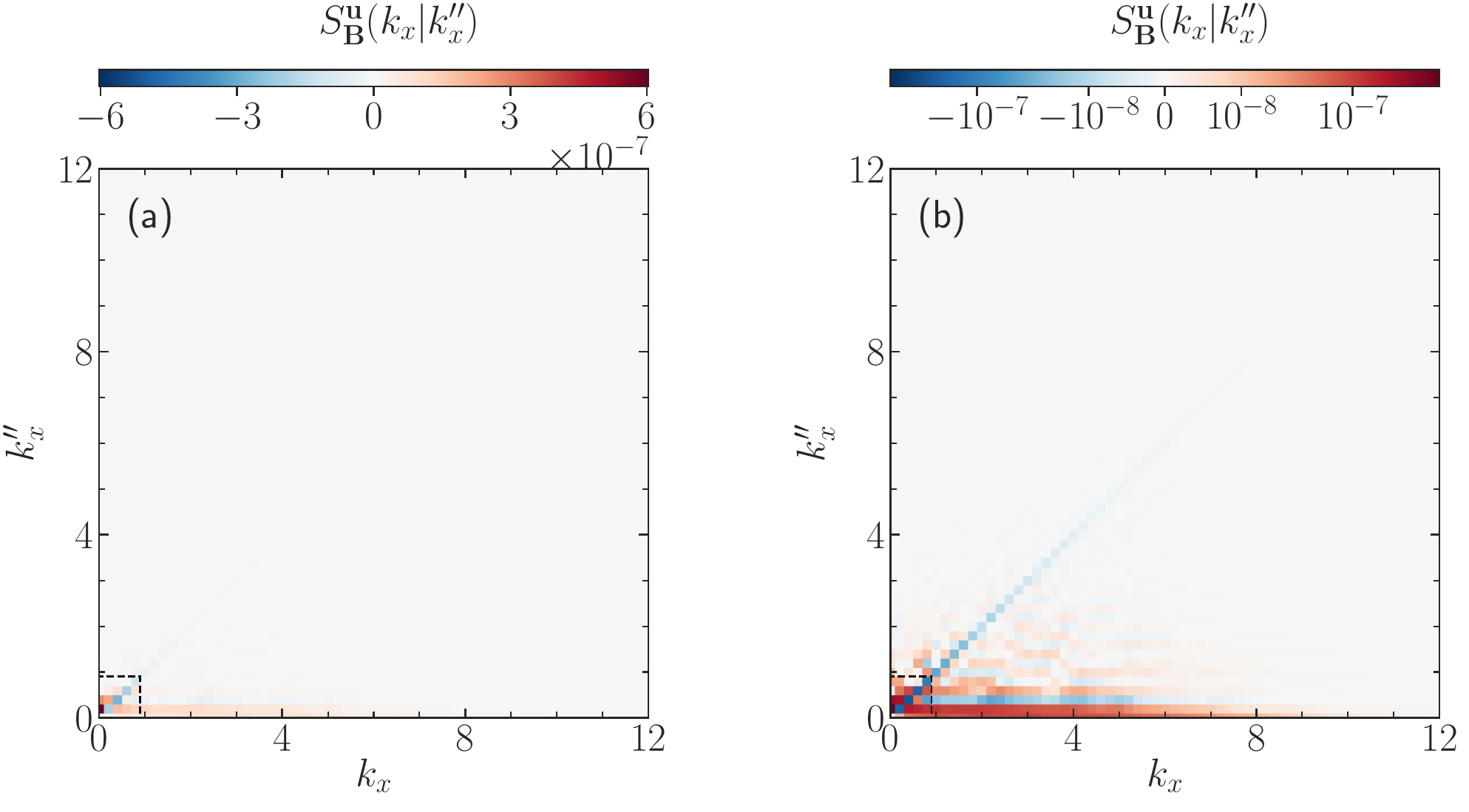}
    \caption{Time-averaged wavenumber-to-wavenumber energy transfer rates between the velocity and magnetic fields. (a) The transfer involves nonlocal triads; the energy transfer is also nonlocal. In particular, the first few Fourier modes (up to around 4) of the flow, shown with $k_x''$, generate significant energy in the magnetic fields even at wavenumbers $k_x$ ranging up to a value, as high as $k_x\sim 8\textrm{--}12$, evidenced in panel (b), using a logarithmic colorbar. The simulation parameters used are $\Ma=120$, $\Dk=2$, and $\Pm=1$.}
    \label{fig:f5}
\end{figure*}

In Fig.~\ref{fig:f2}(b), the nonlinear processes that deposit or remove energy at each scale balance the linear processes. This suggests that the forced turbulence at hand is quasi-stationary in nature.
The nonlinear transfer term combines contributions from all scales in all possible triads that collectively give or take energy away from each scale $k_x$. It does not distinguish whether the energy at $k_x$ comes from the velocity or magnetic field at other wavenumbers of the triad. To identify and quantify such detailed contributions, we compute the wavenumber-to-wavenumber energy transfer functions between and among the velocity and magnetic fields, represented by the four transfer functions $S^\vecu_\vecu(k_x\vert k_x''), S^\vecb_\vecb(k_x\vert k_x''), S^\vecu_\vecb(k_x\vert k_x''), \mathrm{and\ }S^\vecb_\vecu(k_x\vert k_x'')$. Recalling that $S^\vecu_\vecb(k_x\vert k_x'') = -S^\vecb_\vecu(k_x''\vert k_x)$, there are only three unique transfer functions. These unique transfer functions are displayed in Figs.~\ref{fig:f3}--\ref{fig:f5}.

In Fig.~\ref{fig:f3}(a), it is seen that, outside the KH-unstable range (shown as a dashed black box in the bottom left corner), the $\vecu$-to-$\vecu$ transfer is almost negligible. Only on using a logarithmic scale of the transfer, an interesting feature is observed---a nonlocal triad, with a local energy transfer.  This aspect is much more active and prominent in $\vecb$-to-$\vecb$ energy transfer in Fig.~\ref{fig:f4}, where a linear scale of energy transfer alone shows such a behavior. The magnetic energy is dominantly transferred to smaller scales from larger scales in an iterative manner.  The iteration occurs in such a way that a Fourier mode number $n$ receives magnetic energy from the mode $n-1$ and gives magnetic energy to $n+1$---as evidenced in the bidiagonal structure in the transfer function in Fig.~\ref{fig:f4}. This also implies that the $\vecb$-to-$\vecb$ energy transfer is local, but the triad involved is highly nonlocal for large $n$.

We observe yet another different feature in the third transfer function in Fig.~\ref{fig:f5}: the flow $\vecu$ at low wavenumbers $|k_x''| \sim 0 \textrm{--} 0.5$  significantly injects energy into the magnetic reservoir at a wide range of wavenumbers.

\begin{figure}
    \centering
    \includegraphics[width=1\textwidth]{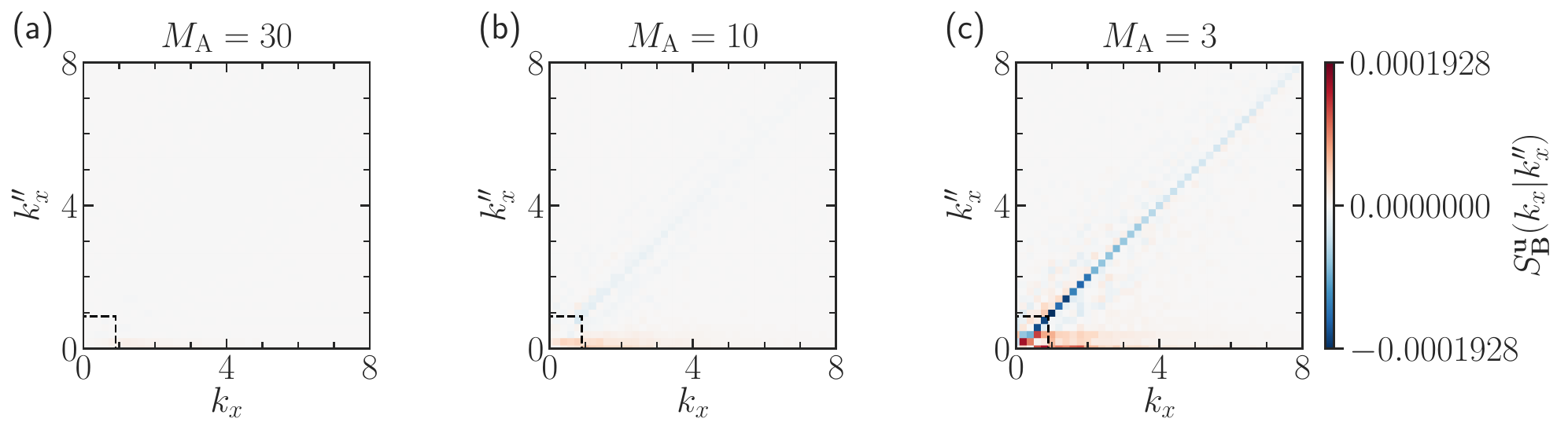}
    \caption{Similar to Fig.~\ref{fig:f5}, but for differing magnetic field strengths. Here, (a) $\Ma=30$, (b) $\Ma=10$, (c) $\Ma=3$. The single diagonal $k_x=k_x''$ appears distinctly in subplot (c), and to a lesser degree in (b). This emerges from $S^{\vecu}_{\vecb}(k_x\vert k_x'') = \real \Big\{ \Big\langle\vecb^\ast(k_x) \cdot \Big[ \vecb(k_x'=0) \cdot \nabla'' \vecu(k_x''=k_x) \Big] \Big\rangle_z \Big\}$, signifying a larger amount of work the small-scales ($|k_x''| \gg 1$) of the velocity fluctuations do, while attempting to bend the mean magnetic-field $\vecb(k_x'\mathrm{=}0)$. 
    Nonlinearly, the first few (up to around $4$) Fourier modes of the flow significantly inject energy into the magnetic fields, even into higher wavenumbers, as discussed in Fig.~3(c). Other simulation parameters used are $\Dk=2$ and $\Pm=1$.}  
    \label{fig:f6}
\end{figure}

The transfer functions are sensitive to the strength of the magnetic field.  When it is very strong ($\Ma \lesssim 10$), a noticeable change is observed in $S^\vecu_\vecb(k_x \vert k_x'')$---the emergence of a diagonal in Fig.~\ref{fig:f6}. Since $\vecu$-to-$\vecb$ transfer is governed by $\real \Big\{ \Big\langle\vecb^\ast(k_x) \cdot \Big[ \vecb(k_x') \cdot \nabla'' \vecu(k_x'') \Big] \Big\rangle_z \Big\}$, and the diagonal of Figure~\ref{fig:f6}(c) implies the transfer occurs from $\vecu(k_x'')$ to $\vecb(k_x)$ with $k_x=k_x''$, it is the mediator field $\vecb(k_x'=0) $ that is responsible for the emergence of the diagonal. The diagonal gets amplified with a stronger magnetic field, and becomes prominent when $\Ma\lesssim 10$. Physically, this effect can be interpreted as the stretching of the stronger mean magnetic-field ($k_x'=0$) by the turbulent flow ($k_x''$) at a wide range of scales, generating larger-amplitude magnetic fluctuations at such scales ($k_x=k_x''$).

\newpage

\subsection{Cross-scale energy fluxes: Forward or inverse cascade?}

\begin{figure*}
    \centering
    \includegraphics[width=1\textwidth]{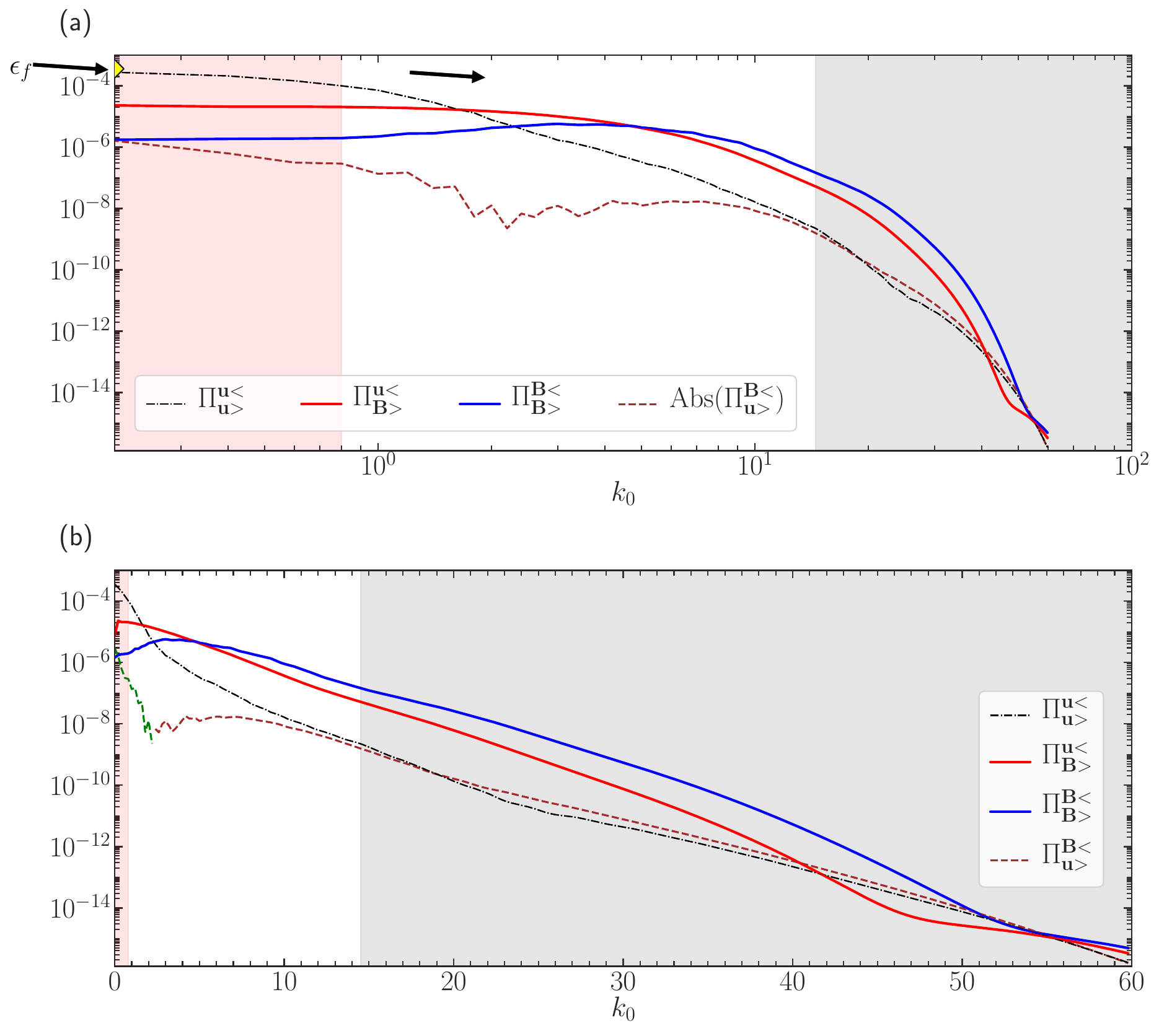}
    \caption{Positive $\Pi$ implies transfer along the arrows of Fig.~\ref{fig:f1}. In the $2$D turbulence here, cross-scale MHD energy fluxes move from large to small scales. (a) The kinetic and magnetic energies both are robustly cascaded forward at every scale, as evidenced by the black and the blue curves, respectively. The kinetic energy flux $\Pi^{\vecu<}_{\vecu>}$ is dominant at the largest scales, and at $k_0=0$ (shown with the yellow diamond on the left axis), the flux agrees with the energy $\epsilon_f = 0.0004$ externally supplied by the Krook forcing  to the mean flow. (b) The fluxes develop a noticeable exponential envelope at around $k_0=15$, after which the exponential fall-off continues with increasing wavenumber. The simulation resolves the dissipation range, as $k_0$ in the simulation ranges up to $205$. The flux $\Pi^{\vecb<}_{\vecu>}$ for $k_0\leq 2.2$ (shown with a green dashed curve) is negative. The grey-shaded region corresponds to the dominant dissipation scales, starting around $k_0=(\epsilon_f\, Re^{3})^{1/4} = 15$; the light-red shaded region is the instability range. Parameters used here are $\Ma=120$, $\Dk=2$, $Re=500$, and $\Pm=1$.}
    \label{fig:f7}
    \end{figure*}

The $\vecb$-to-$\vecb$ transfer in Fig.~\ref{fig:f4} is consistent with 
a forward cascade of magnetic energy.
This is further confirmed from the cross-scale energy fluxes
through a fixed wavenumber $k_0$. Using Eqs.~\eqref{eq:flux1}--\eqref{eq:flux4},  which quantify the energy passing through $k_0$, 
energy transfer from low wavenumbers $|k_x| \leq k_0$ to high wavenumbers $|k_x|>k_0$ can be measured. In Fig.~\ref{fig:f7}, except at the lowest wavenumbers $|k_0| \lesssim 2$, near the KH-instability range, all energy fluxes are robustly in the forward direction---along the arrows shown in Fig.~\ref{fig:f1}. The forward cascades of energies are clearly observed in Fig.~\ref{fig:f7}(b).
Note that, at most of the scales, $\Pi^{\vecb\mathrm{<}}_{\vecb\mathrm{>}}$ is larger than $\Pi^{\vecu\mathrm{<}}_{\vecb\mathrm{>}}$, which is in turn larger than $\Pi^{\vecu\mathrm{<}}_{\vecu\mathrm{>}}$, in agreement with the energy transfer in every triad pair shown in Figs.~\ref{fig:f3}\textrm{--}\ref{fig:f5}.  The forward cascade of energy (magnetic and kinetic) is expected for $2$D MHD \cite{biskamp2003}, and is intrinsic to the fluxes of Eqs.~\eqref{eq:flux1}--\eqref{eq:flux4}, which are based on energy transfer rates.  

Inverse cascading in $2$D homogeneous MHD applies only to the mean-squared magnetic-flux-function, $\int \psi^2 d^2x$, whose corresponding nonlinear transfer carries fewer spatial derivatives than the nonlinear transfer of energy. In this paper, we are expressly interested in the energetics of the inhomogeneous-flow-driven turbulence.

\section{Energetic coupling of eigenmodes} \label{sec:sec5}
\subsection{Eigenmode interaction with evolving mean profiles}\label{sec:sec5}

Insights into instability saturation can be obtained by tracing the interaction of the unstable and stable eigenmodes with the mean profiles. Since the instantaneous mean flow profile, although forced, deviates from the initial mean-flow, there is, in general, some linear coupling between the linear eigenmodes of the initial mean profile. Whether this \mbox{(quasi-)}linear coupling is comparable to the linear growth rate of the eigenmodes of the initial profile is an important question. If the energy exchange rate $R_j$ between the fluctuations and the time-deviation of the mean profiles via this \mbox{(quasi-)}linear coupling is small compared to the energy exchange rate $Q_j$ between the initial mean profiles and the fluctuations, then using the eigenmodes of the initial mean profiles is justifiable for understanding energy dynamics and nonlinear coupling. 
To assess this exchange, we first compute the expressions in Eqs.~\eqref{eq:self} and \eqref{eq:cross} and compare them as time evoles for both the unstable ($j=1$) and stable ($j=2$) eigenmodes. 

\begin{figure*}
    \centering
    \includegraphics[width=0.99\textwidth]{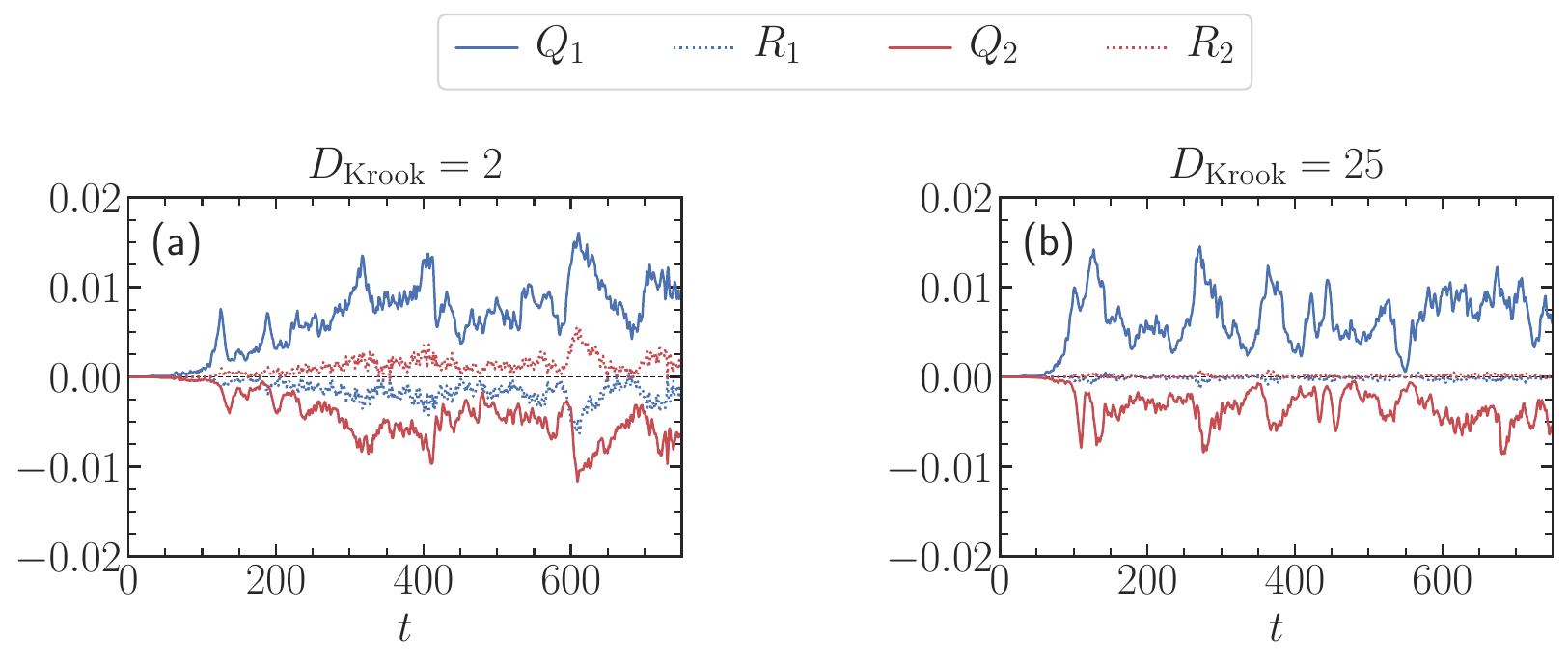}
    \caption{Linear processes of energy transfer due to unstable ($j=1$) and stable ($j=2$) eigenmodes. Shown are the linear injection or withdrawal rates $Q_j$ using the profiles at $t=0$, and the rate $R_j$ using deviations of the instantaneous mean profiles of the flow and the magnetic field from their initial profiles. The wavenumber of the fluctuation chosen is $k_x=0.2$, which is the loweest non-zero wavenumber in the simulation, where both the linear drive rate as well as the  fluctuation energy spectrum peak. 
    Comparing (a) $\Dk=2$ with (b) $\Dk=25$ shows that the higher rate (i.e., the larger $\Dk$) of replenishment of the mean flow via Krook forcing removes almost all linear coupling---the term $R_j$---that is induced due to instantaneous fluctuations in the mean profiles. The simulation parameters used are $\Ma=10$ and $\Pm=1$.}
    \label{fig:f8}
\end{figure*}

The rate at which the external forcing replenishes the mean flow can control the deviations of the instantaneous mean profiles from the initial profiles. We investigate this aspect in Fig.~\ref{fig:f8}. Both cases of forcing strengths show that the effect of the linear coupling between the eigenmodes---quantified by $R_j$---is considerably smaller than $Q_j$. The ratio of $R_j$ to $Q_j$ gets further lowered when the mean flow is replenished faster (i.e., larger $\Dk$).  This means that the evolved shear-flow profile is closer to the initial unstable flow-profile. This detailed analysis provides justification for using eigenmodes of the initial profile as a fluctuation basis.

We note that the energy transfer $Q_2$ is appreciable whenever the stable modes are excited to large amplitudes.  Their excitation at early stage is caused by the nonlinear interaction between the unstable modes at different wavenumbers.\cite{terry2006} So, whenever there are at least a few Kelvin-Helmholtz unstable wavenumbers,\cite{fraser2017} stable modes are universally, significantly excited; $Q_2$ then becomes comparable to $Q_1$.

\begin{figure}
    \centering
    \includegraphics[width=0.7\textwidth]{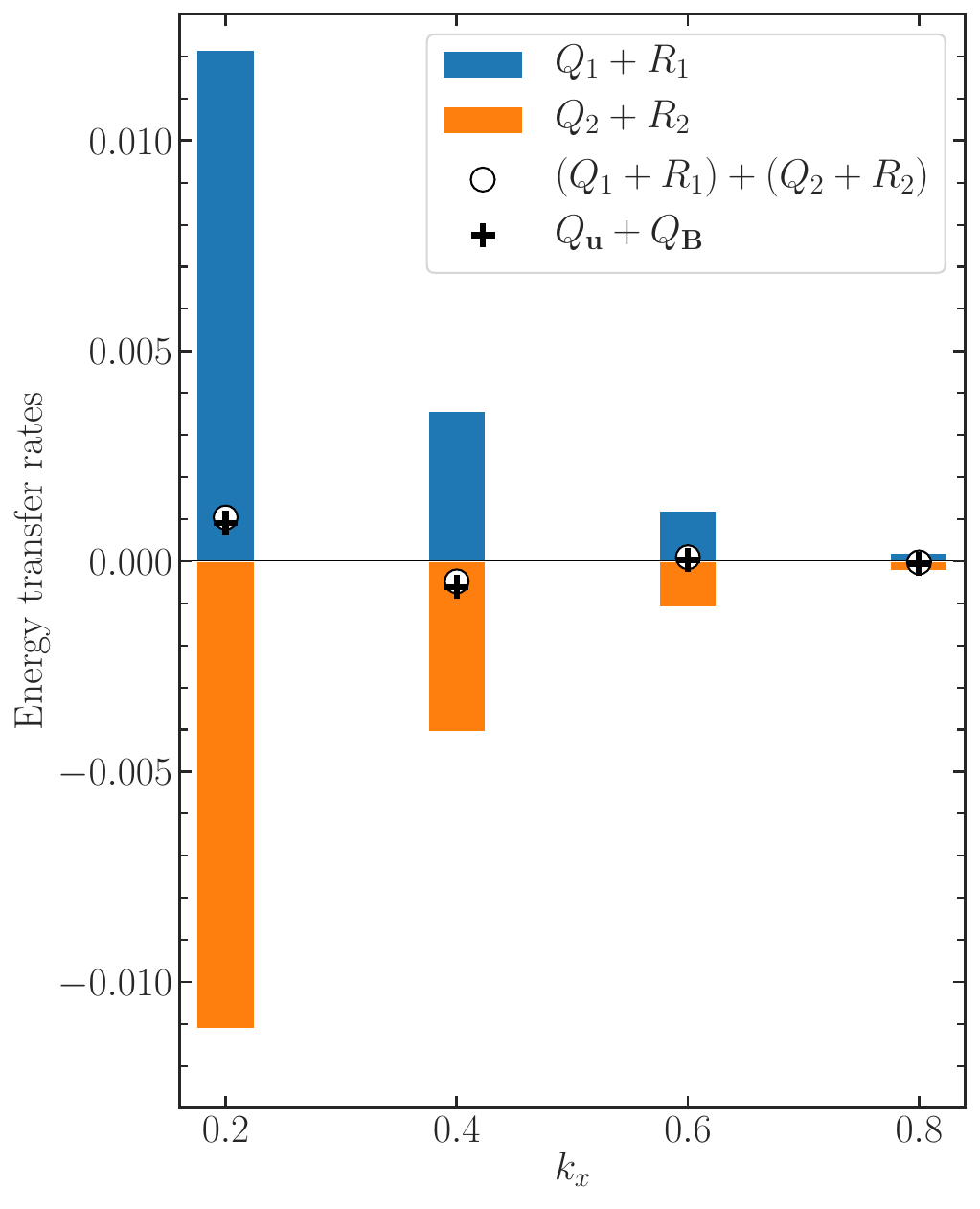}
    \caption{Comparison of time-averaged energy transfer rates $Q_j+R_j$ from the instantaneous mean profiles of the flow and the magnetic fields for the unstable ($j=1$) with that for the stable ($j=2$) modes. These rates are then summed to predict the total energy transfer rates via linear processes, which also include the contributions from the continuum eigenmodes and the small visco-resistive dissipation occurring at these large scales. The predictions, shown with black circles, are in excellent agreement with the total linear energy transfers, depicted with black plus signs.}
    \label{fig:f9}
\end{figure}

Now, we use the eigenmode decomposition to investigate why some wavenumbers, despite lying in the KH-unstable wavenumber range (in Fig.~\ref{fig:f2}), withdraw energy from the fluctuation spectrum instead of depositing.
In Fig.~\ref{fig:f9}, we show the energy transfer via linear processes, 
decomposing it into the contribution from unstable and stable modes. To show how well the sum of the contributions from unstable and stable modes 
captures the total $z$-integrated transfer, we overlay the data from Fig.~\ref{fig:f2}. 
There, contributions from the continuum eigenmodes are also included.  
At $k_x=0.2$,  it is clearly observed that the stable mode transfers energy at a slightly greater rate than
the unstable modes, thus depleting the overall fluctuation energy at that wavenumber. The quantity $Q_1+Q_2$ captures reasonably well the total energy transfer $Q_\vecu+Q_\vecb$. Any discrepancy is attributed to the contributions from the sum of all the continuum modes, which dominates over the small visco-resistive dissipation of the unstable and stable modes at these large length scales.

\newpage

\subsection{Channels of energy transfer between eigenmodes}\label{sec:sec6}

\begin{figure}
    \centering
    \includegraphics[width=0.95\textwidth]{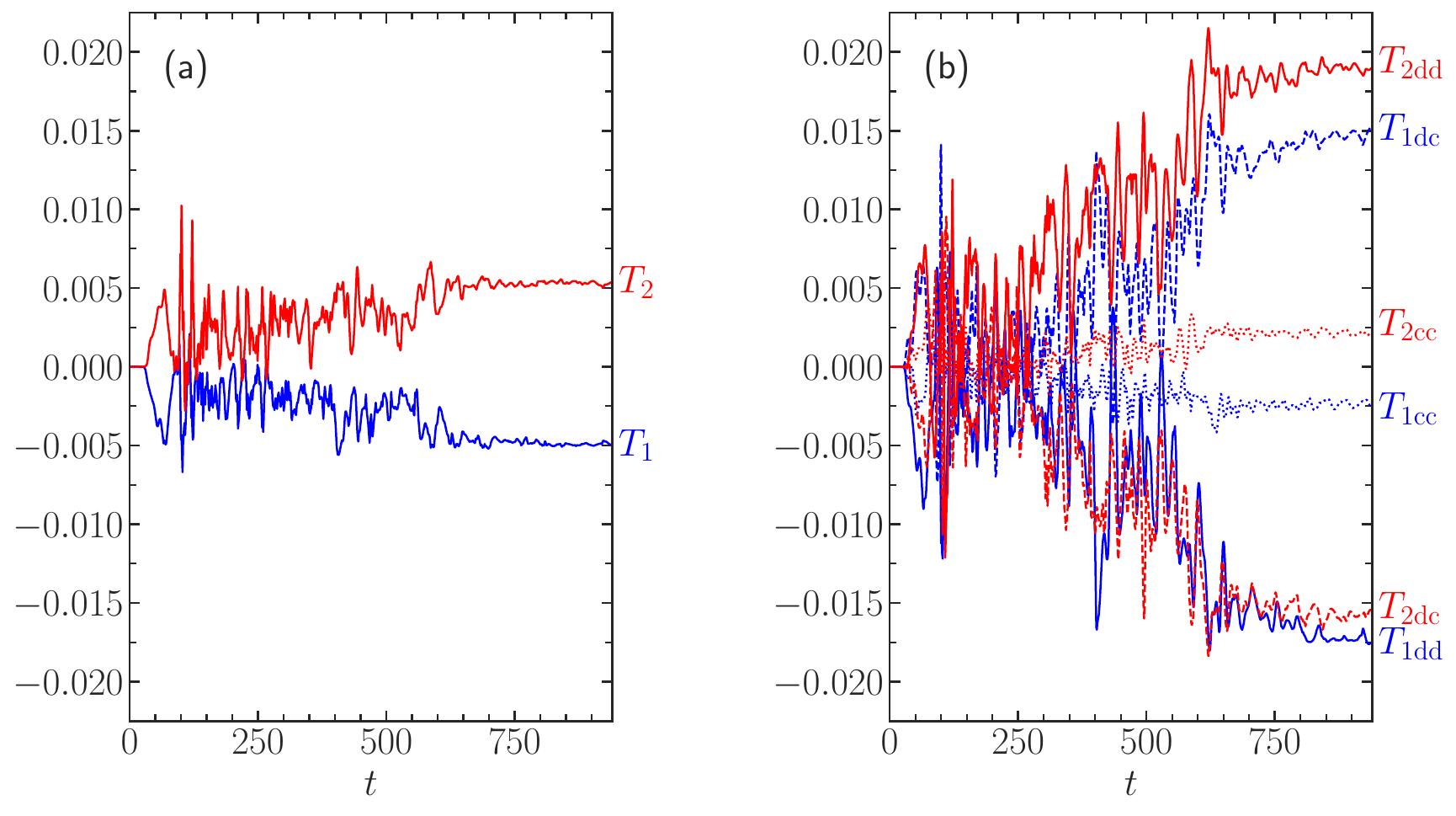}
    \caption{(a) Nonlinear energy transfer $T_{1}(k_x)$ that saturates unstable modes and $T_{2}(k_x)$ that excites stable modes. (b) These rates are then decomposed into three classes of triadic interactions, where the beating modes are discrete-discrete ($T_{j\mathrm{dd}}$), or discrete-continuum ($T_{j\mathrm{dc}}$), or continuum-continuum ($T_{j\mathrm{cc}}$). The $T_{j\mathrm{cc}}$ are negligibly small. The simulation parameters are $\Pm=0.1, \Ma=10,$ and $\Dk=2$, and the chosen wavenumber is $k_x=0.4$, which is the linearly fastest growing wavenumber.}
    \label{fig:f10}
\end{figure}

We 
now analyze the nonlinear excitation and saturation processes for stable and unstable modes. 
Figure~\ref{fig:f10} shows the terms that saturate (take away energy from) the unstable modes $T_1$ and drive (feed energy into) the stable modes $T_2$. Near-equal levels of $T_1$ and $T_2$ are observed. These transfer terms are further probed in Fig.~\ref{fig:f10}(b), where three sub-classes of triadic interactions are shown---$T_{j\mathrm{dd}}, T_{j\mathrm{dc}}$, and $T_{j\mathrm{cc}}$, representing the nonlinear interactions among the discrete modes, between the discrete and continuum modes, and among the continuum modes, respectively. These three terms capture all the nonlinear terms appearing on the right-hand side of the evolution equation of the $j^\mathrm{th}$ mode, with $j=1$ and $j=2$ representing the unstable and stable modes, respectively. We compute these three nonlinear terms following the expressions in Eqs.~\eqref{eq:Tjdd}--\eqref{eq:Tjcc}. We find that $T_{j\mathrm{cc}}$ is appreciably smaller than $T_{j\mathrm{dd}}$ and $T_{j\mathrm{dc}}$, which are of similar magnitude but of opposite sign. To be precise, $|T_{j\mathrm{cc}}| \ll |T_{j\mathrm{dc}}| \lesssim |T_{j\mathrm{dd}}|$.

\begin{figure}
    \centering
    \includegraphics[width=0.98\textwidth]{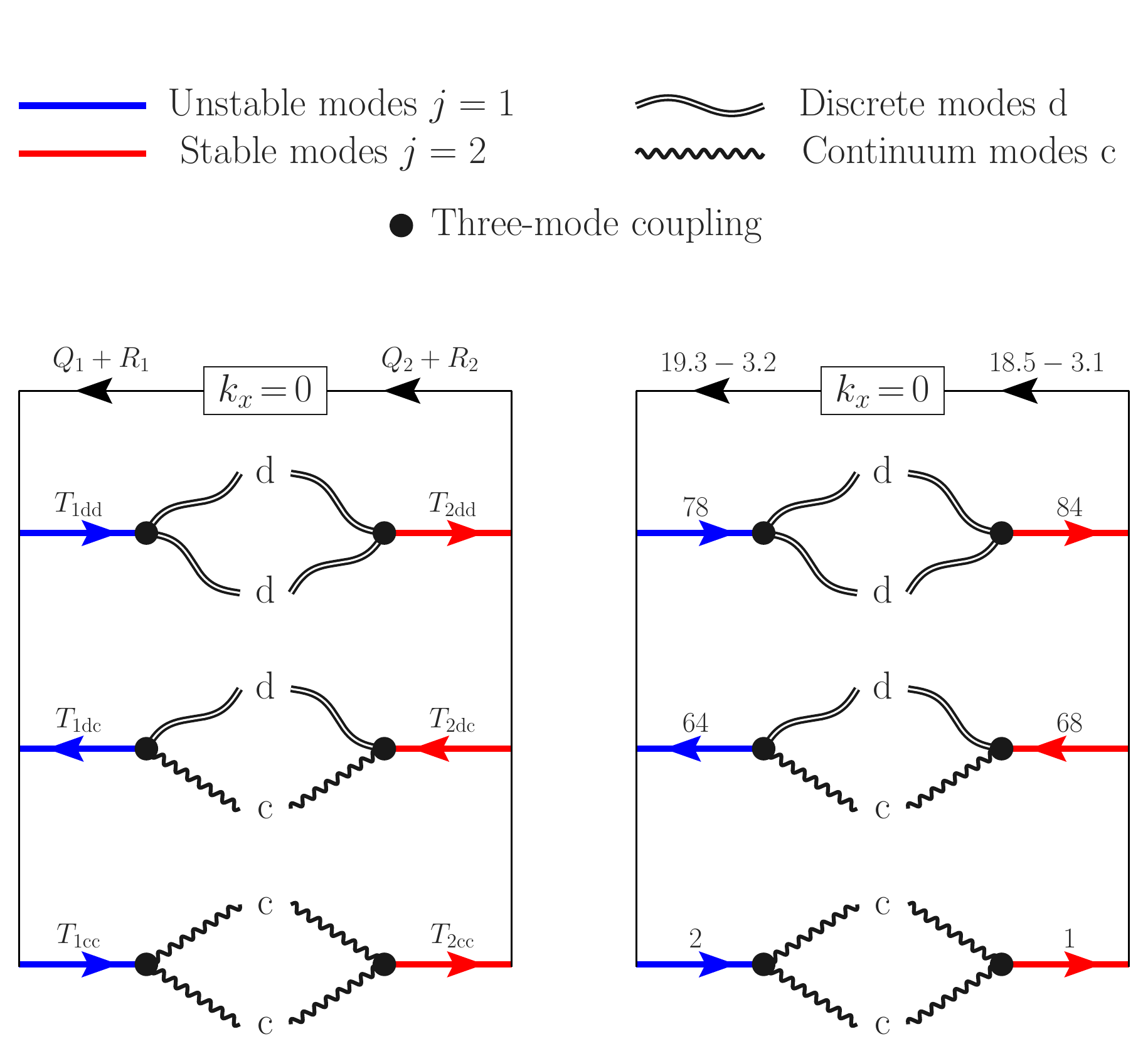}
    \caption{(Left) A schematic diagram that shows the terms corresponding to both the linear and nonlinear energy transfer processes that drive and saturate the unstable and stable modes. Since $Q_j, R_j, T_{j\mathrm{dd}}, T_{j\mathrm{dc}},$ and $T_{j\mathrm{c}}$ all appear on the right hand-side of Eq.~\eqref{eq:dtbeta2}, it is expected, on time-averaging, that $|Q_j+R_j| \approx |T_{j\mathrm{dd}} + T_{j\mathrm{dc}} + T_{j\mathrm{c}}|$ for both $j=1$ and $j=2$, individually. (Right) The energy transfer terms of the left-hand schematic diagram is quantified in units of $10^{-3} U_0^3/a$, for a simulation with parameters $\Pm=0.1, \Ma=10,$ and $\Dk=2$. The difference between $Q_1+R_1$ and $Q_2+R_2$ represents the rate of energy that is supplied by the Krook forcing to the mean flow, attempting to restore the original flow-profile.}
    \label{fig:f11}
\end{figure}

\begin{figure}
    \centering
    \includegraphics[width=0.98\textwidth]{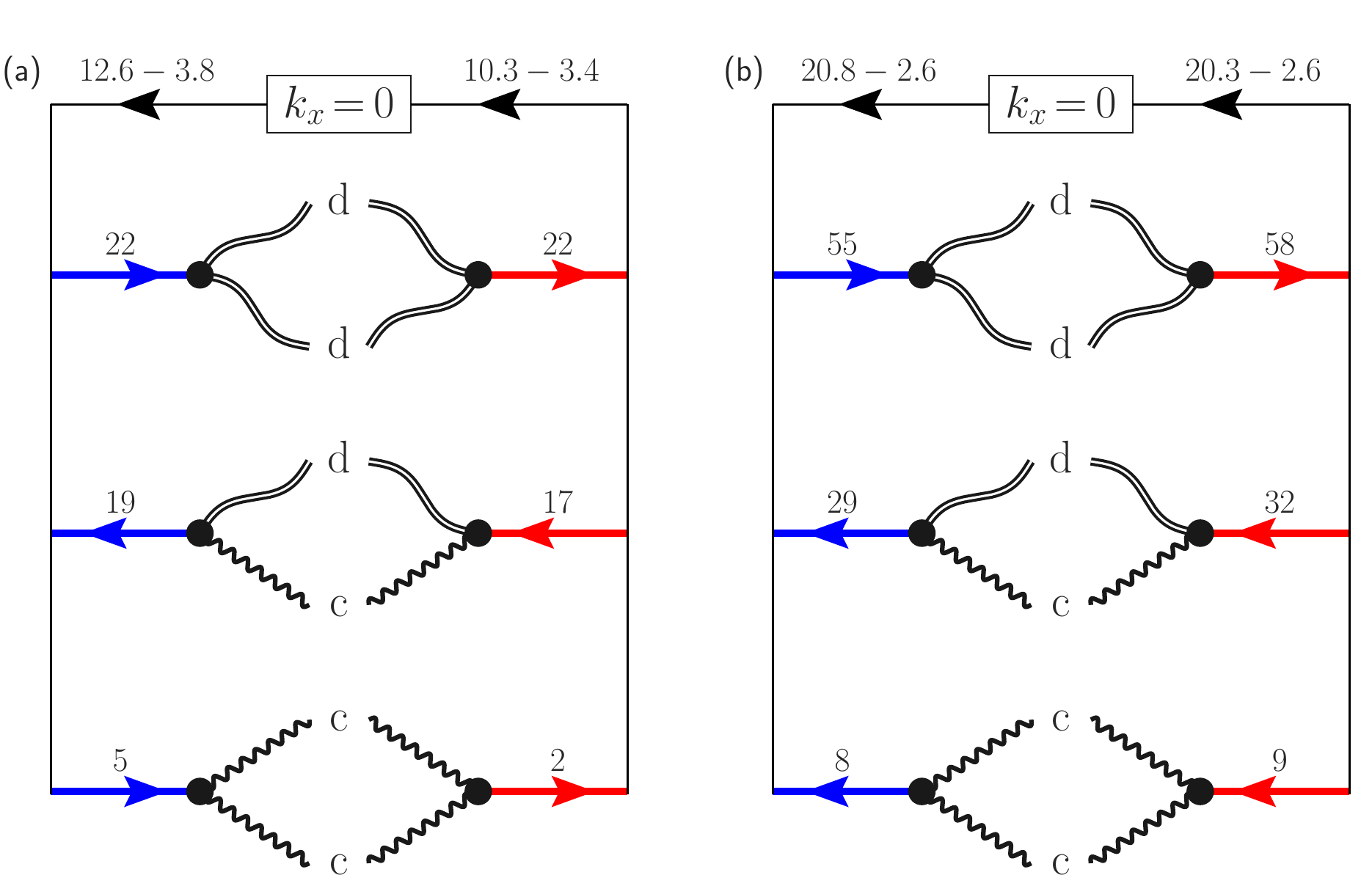}
    \caption{Same as the right-hand panel in Fig.~\ref{fig:f11}, but now with varying magnetic field strength---the magnetic Prandlt number is also changed to $\Pm=1$. The parameters used are (a) $\Ma=10$ and (b) $\Ma=120$, both with the forcing strength $\Dk=2$. Energy transfer rates are measured in units of $10^{-3} U_0^3/a$. With weaker magnetic field, Lorentz feedback on the flow of unstable and stable modes is weakened, and hence discrete-discrete interaction dominates over the discrete-continuum interaction.} 
    \label{fig:f12}
\end{figure}

We now investigate whether the above apportionment of the  three kinds of nonlinear transfer is intrinsic to a particular simulation-parameter regime or whether magnetic fields and magnetic Prandtl number (the ratio of viscosity to resistivity) affect these channels. We plot time-averaged transfer rates, both linear and nonlinear, in a single diagram. In Fig.~\ref{fig:f11}(a), we first show the schematic diagram and then annotate it with numerical time-averaged transfer rates in Fig.~\ref{fig:f11}(b), where $\Pm=0.1$. Indeed, $T_{j\mathrm{cc}}$ is negligibly small compared to $T_{j\mathrm{dc}}$, which is only a little smaller than $T_{j\mathrm{dd}}$, even for $\Pm=1$ in Fig.~\ref{fig:f12}(a). Weakening the magnetic field strength in Fig.~\ref{fig:f12}(b) does not alter $T_{j\mathrm{cc}}$ dramatically, but $T_{j\mathrm{dc}}$ does become appreciably smaller than $T_{j\mathrm{dd}}$. We interpret this as a result of the lower level of fluctuations of the continuum modes, which have been found to capture the magnetic fluctuations\cite{tripathi2022b}; the magnetic fluctuations are weaker when the mean magnetic field is weak. In this case, the discrete modes of the flow assume a dominant role in nonlinearly driving the stable modes.

\newpage

\subsection{Predicting energy transfer from mode-coupling coefficients}

Statistical closure theories of turbulence\cite{orszag1970, terry2018} predict that the mode-coupling coefficient is a key factor, although not necessarily the only factor; mode energy levels and the three-wave correlation time also enter the formula for the nonlinear energy transfer rate. \cite{makwana2012} 

\begin{figure}
    \centering
    \vspace{1cm}\includegraphics[width=0.85\textwidth]{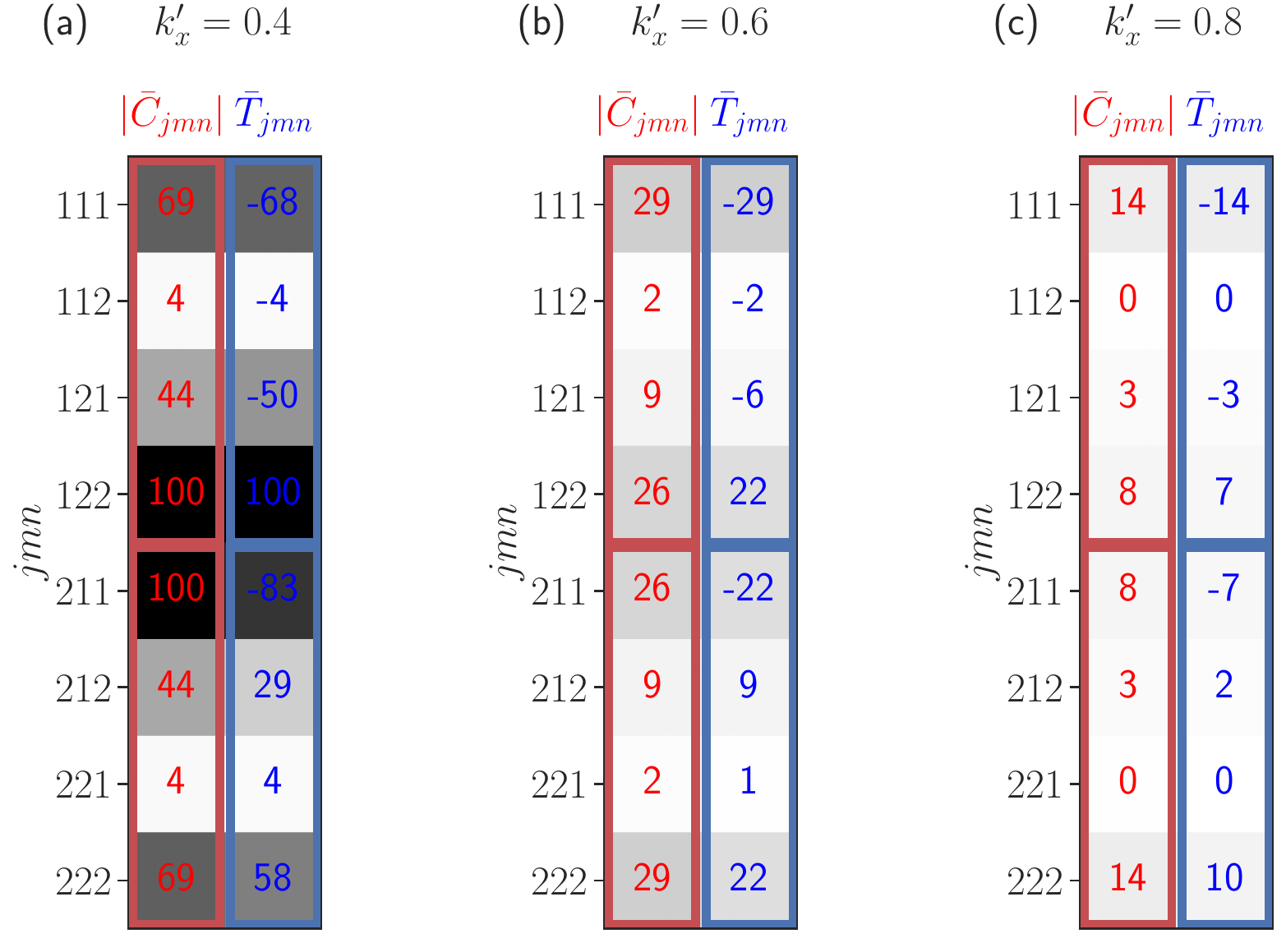}
    \caption{Comparison of the symmetrized nonlinear mode-coupling coefficient $\bar{C}_{jmn}(k_x, k_x') $ and the symmetrized nonlinear energy transfer $\bar{T}_{jmn}(k_x, k_x')$. The transfer term represents energy being pumped into an eigenmode $j$ at wavenumber $k_x=0.2$ due to nonlinear interaction between eigenmodes $m$ at $k_x'$ and $n$ at $k_x''=k_x-k_x'$. The largest nonlinear drive is at $k_x=0.2$ (the first Fourier mode). In the chart, all $48$ possible $\mathrm{discrete}\textrm{-}\mathrm{discrete}$ interactions, decomposed by individual unstable and stable modes and labeled by $jmn$, are shown, for (a) $k_x'=0.4$, (b) $k_x'=0.6$, and (c) $k_x'=0.8$. In each subplot, the coupling coefficients have been scaled by the same factor.  The transfer rates are measured in units of $10^{-3} U_0^3/a$. Positive (negative) transfer feeds (withdraws) energy in (from) an eigenmode $j$ at $k_x=0.2$. The simulation parameters used are $\Ma=10$ and $\Dk=2$. Note the symmetry reflected in the upper and lower 4 rows within each subplot, and the strong correlation between the coupling coefficient and the energy transfer.}
    \label{fig:f13}
\end{figure}

To learn if the nonlinear mode-coupling coefficients are predictive of nonlinear energy transfer between eigenmodes in the quasi-stationary state of turbulence, we now separate the nonlinear energy transfer $T_{j\mathrm{dd}}$ further into individual components $T_{jmn}$ by decomposing the discrete modes into unstable and stable modes: $T_{j\mathrm{dd}} = \sum_{m=1}^2 \sum_{n=1}^2 T_{jmn}$, where $m=1$ stands for unstable modes and $m=2$ represents stable modes, and likewise for $n$. The nonlinear mode-coupling coefficients $C_{jmn}(k_x, k_x')$ of a mode $j$ at $k_x$, nonlinearly coupled with a mode $m$ at $k_x'$ and $n$ at $k_x''=k_x-k_x'$, can be computed. Since the mode-coupling coefficient $C_{jmn}(k_x, k_x')$ is a complex-valued quantity, we compare their absolute values and predict the energy transfer levels $|T_{jmn}(k_x, k_x')|$. To reduce the number of possible nonlinear mode-coupling terms, we now compose a symmetrized coupling coefficient $\bar{C}_{jmn}(k_x, k_x')$ and a symmetrized nonlinear energy transfer $\bar{T}_{jmn}(k_x, k_x')$, and display them in Fig.~\ref{fig:f13} --- where $\bar{C}_{jmn}(k_x, k_x') = C_{jmn}(k_x, k_x') + C_{jnm}(k_x, k_x'')$ and $\bar{T}_{jmn}(k_x, k_x') = T_{jmn}(k_x, k_x') + T_{jnm}(k_x, k_x'')$. 

Strong correlation between the coupling coefficient and the energy transfer suggests that the coupling coefficients are critical elements in setting the energy transfer. This property may enable the construction of reduced-order models for nonlinear saturation based on these coupling coefficients.

\section{Testing a general quasilinear theory of instability saturation}\label{sec:sec7}

\begin{figure}[htbp!]
	\includegraphics[width=0.53\textwidth]{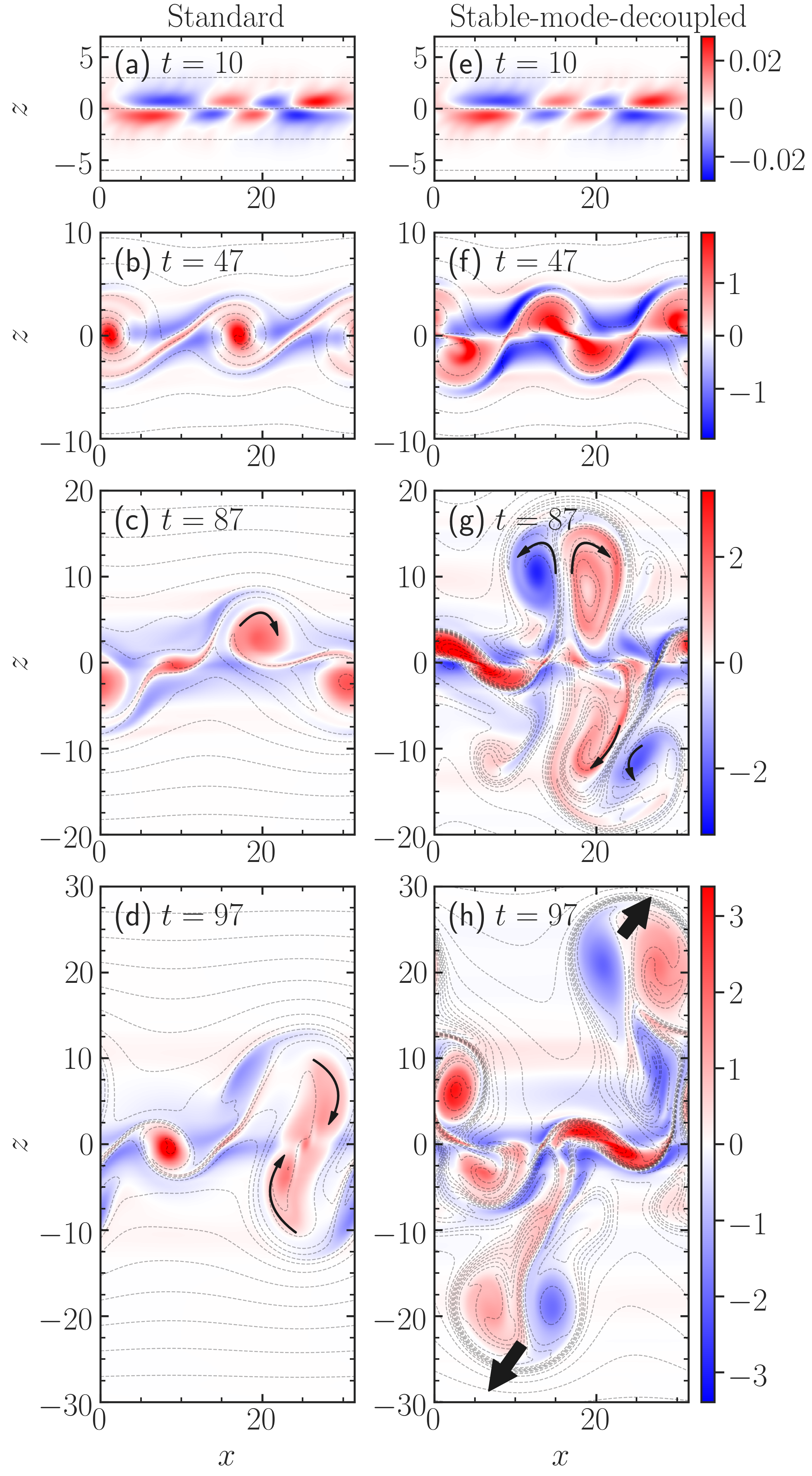}
	\caption{Time evolution of vorticity (filled contour) in two different simulations: (a)--(d) ``Standard," and (e)--(h) ``Stable-mode-decoupled."  The grey, dashed contours show the total magnetic flux function $\psi$, along which the magnetic field lines are aligned. The fields are initially oriented along the positive $x$-axis. In all panels, same number of contours is shown. Unstable modes dominate at the early stage; even in the nonlinear phase, the unstable modes are qualitatively seen in (f)--(h), conspicuously in (f). Thin curved arrows illustrate eddy motions in (c), (d), and (g). Vortices merge in (d), whereas they separate in (h) with a violent ejection of eddies away from the shear layer, shown with thick straight arrows. In (a)--(d), however, stable modes confine the turbulence near the shear layer. Magnetic structures in (g) and (h) are highly folded. (Multimedia view)}\label{fig:f14}
\end{figure}

\begin{figure*}
    \centering
    \includegraphics[width=1\textwidth]{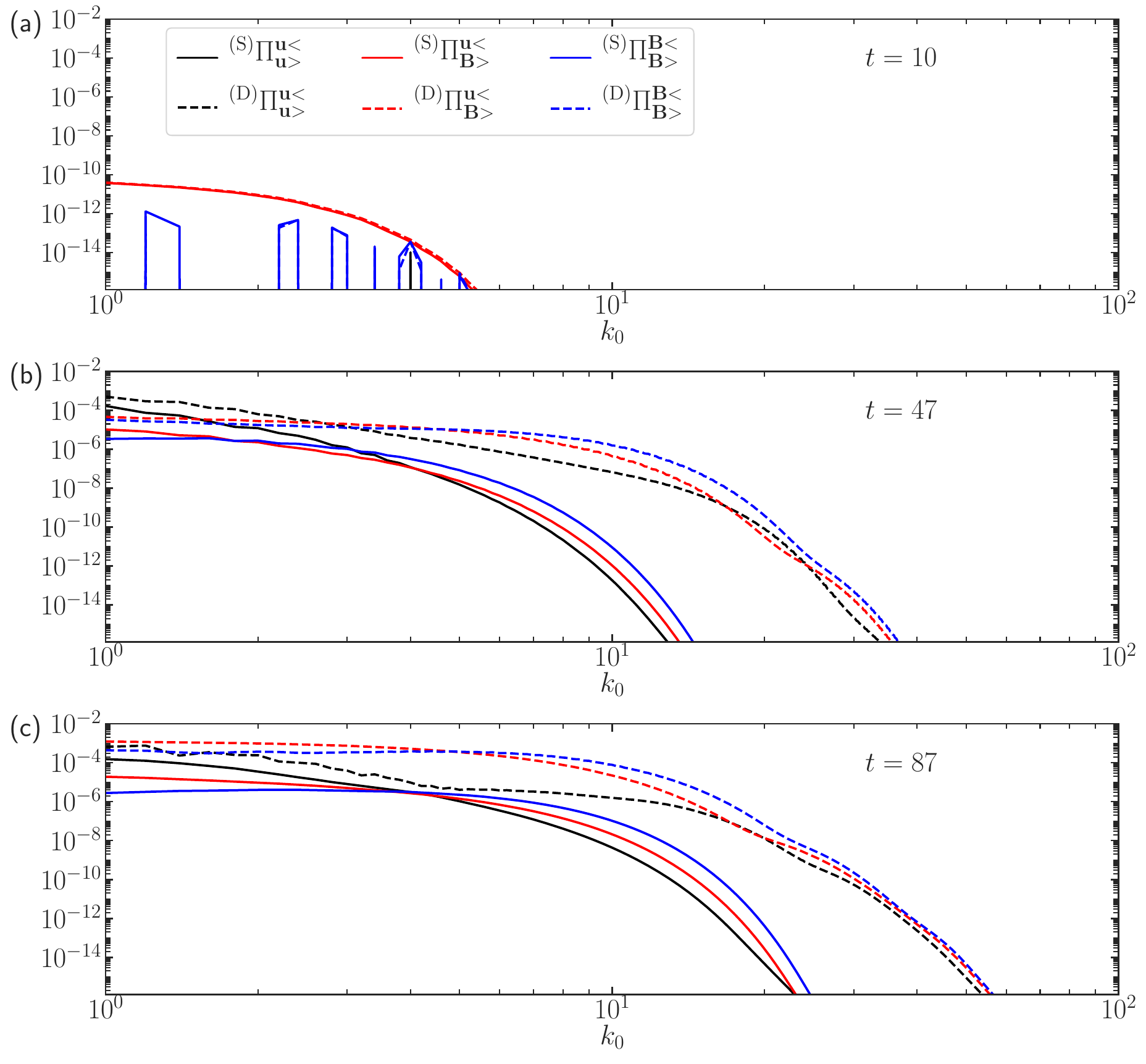}
    \caption{Energy flux shown at wavenumbers, beyond the instability scale $0<k_0<1$. The left-handed superscripts ``(S)" and ``(D)" represent the energy fluxes in a \textit{s}tandard simulation and in a stable-mode-\textit{d}ecoupled simulation, respectively. Both simulations use identical parameters: $\Ma=60$, $\Dk=2$, and $Pm=1$. At the first few time steps, the flux spectra are alike, but soon evolve differently, c.f. (a) with (b). Positive fluxes mean nonlinear energy cascade to small scales. The flux spectra at $t=97$ are similar to that in (c). When uninhibited by stable modes, the energy fluxes are larger by order of magnitude, and the dissipation length scale is pushed to larger $k_0$.}
    \label{fig:f15}
\end{figure*}

At the heart of all quasilinear theories, one presumes that only a particular family of eigenmodes, most often the unstable mode branch, interacts among itself at different scales leading to an instability saturation. Various approximations then are taken to derive simpler forms of quasilinear theories.  Without using any approximation, we wish to test here a general quasilinear model of Kelvin-Helmholtz-instability-driven turbulence where Eqs.~\eqref{eq:mhdu}--\eqref{eq:divb} will be kept fully intact, except that the system will be conditioned not to couple with the large-scale ($|k_x|<1$) conjugate-stable modes. Noting the prevailing notion of instability saturation,\cite{fuller2019, pessah2006, goodman1994, garaud2018} one may assume that the instability-driven nonlinear system, devoid of the stable modes, would produce essentially the same results as one obtains from a standard numerical solution of Eqs.~\eqref{eq:mhdu}--\eqref{eq:divb}. Is that really true?

To answer and test such a general model of instability saturation, Eqs.~\eqref{eq:mhdu}--\eqref{eq:divb} can be transformed to the eigenmode basis of the linear operator. To do so, we follow the same method as described earlier in arriving at Eq.~\eqref{eq:timematured1} from Eq.~\eqref{eq:structuraleq}. Note that the eigenmodes of the non-dissipative linear operator of the shear-flow instability has an unstable, a stable, and a (theoretically infinite) number of continuum modes at each wavenumber in the range $0 < |k_x| < 1$. For $|k_x| \geq 1$, only continuum modes exist. Since the continuum modes are numerous and they are not the ideal choice for a basis function to implement in a numerical solver,
we develop here a novel method to time-evolve the sum of all continuum-mode-associated fluctuations. First, we decompose an instantaneous state vector $X = [\phi, \psi]$ into an $x$-averaged mean $X(k_x=0)$ and fluctuations $\widetilde{X}(x,z)$. The fluctuations are then decomposed as
\begin{equation} \label{eq:continuumplusdiscrete}
\widetilde{X}(x,z) = \sum_{0 < |k_x| < 1}\sum_{j=1}^{2} \beta_j(k_x) X_j(k_x,z) \mathrm{e}^{i k_x x} + X_\mathrm{c}(x,z),    
\end{equation}
where the first term on the right-hand side is a sum of the unstable ($j=1$) and stable modes ($j=2$) throughout the Kelvin-Helmholtz-unstable wavenumber range; and the second term $X_\mathrm{c}(x,z)$ stands for all remaining fluctuations, composed of continuum modes, whose evolution equation can be derived with the help of Eq.~\eqref{eq:continuumplusdiscrete} and is given by
\begin{equation} \label{eq:continuumplusdiscrete2}
\begin{aligned}[b]
\partial_t M X_\mathrm{c}(x,z) 
&= \partial_t M  \left[\widetilde{X}(x,z)-\sum_{0 < |k_x| < 1}\sum_{j=1}^{2} \beta_j(k_x) X_j(k_x,z) \mathrm{e}^{i k_x x}\right]\\
&= \partial_t M  \widetilde{X}(x,z) - M  \sum_{0 < |k_x| < 1}\sum_{j=1}^{2} \partial_t \beta_j(k_x) X_j(k_x,z) \mathrm{e}^{i k_x x},
\end{aligned}
\end{equation} 
where the linear operator $M$ is defined in Eq.~\eqref{eq:mop}.

The second equality of Eq.~\eqref{eq:continuumplusdiscrete2} can be explicitly expressed using Eq.~\eqref{eq:structuraleq} as
\begin{equation} \label{eq:allcont}
    \partial_t M X_\mathrm{c}(x,z) =L_0 \widetilde{X} + \widetilde{L} \widetilde{X} + L_\mathrm{diss} \widetilde{X} + N (\widetilde{X}, \widetilde{X}) 
    - M  \sum_{0 < |k_x| < 1}\sum_{j=1}^{2} \partial_t \beta_j(k_x) X_j(k_x,z) \mathrm{e}^{i k_x x},
\end{equation}   
and $\partial_t \beta_j$ in Eq.~\eqref{eq:allcont} can be replaced with the right-hand side of Eq.~\eqref{eq:betaevoln1}, which is repeated below for convenience,
\begin{equation}\label{eq:dtbeta12decoupling}
    \partial_t  \beta_j(k_x) = \gamma_j(k_x) \beta_j(k_x) + \left\langle Y_j, \left[\widetilde{L}  \hat{X} + L_\mathrm{diss} \hat{X} \right] \right\rangle+ \sum_{k_x'} \left\langle Y_j, \hat{N} (\hat{X}', \hat{X}'') \right\rangle \ \ \ \ \mathrm{for\ }j=1,2.
\end{equation}

In a numerical simulation, termed ``Standard", we evolve $k_x=0$ mode in Eq.~\eqref{eq:phispsieqn}. This equation, however, couples with the fluctuations; hence, Eq.~\eqref{eq:betaevoln1}, with $j=1,2$, and Eq.~\eqref{eq:allcont} are solved in conjunction. Such a solution reproduces the solution obtained from the usual direct numerical simulation of Eq.~\eqref{eq:phispsieqn} to machine precision. This is anticipated, as Eq.~\eqref{eq:betaevoln1} is obtained from merely a change of basis.

In another simulation, termed ``Stable-mode-decoupled," two changes are made to the ``Standard" system. First, Eq.~\eqref{eq:betaevoln1} with $j=2$ is erased.  Second, from the entire system, all terms containing $\beta_2$ are removed---that is, coupling to the stable modes is analytically removed from the nonlinear system. Removing a family of eigenmodes in such a careful way by hand, rather than numerically zeroing them out at each time step of evolution, is unconventional, but has been applied in a few other cases, e.g., removal of a family of helical modes in isotropic and homogeneous turbulence.~\cite{biferale2012}  Since the modes are completely removed from the system, they neither receive nor donate energy in a triadic interaction, and hence the mode-removed equations conserve the ideal invariants of the full standard nonlinear system.

The solutions of vorticity from the above two numerical simulations are visualized in Fig.~\ref{fig:f14} (Multimedia view). Both simulations have identical parameters $\Ma=60$, $\Dk=2$, and $Re=Rm=50$.
When the stable modes are analytically removed, the figure shows that the turbulence reaches higher amplitudes and becomes violent. In this case, the unstable modes can saturate only by passing on their energy to the small-scale cascade, which involves the generation of extended secondary flow structures. 
When stable modes are kept intact in the equations, they confine the turbulence near the shear layer, and thus lead to vortex merging events, as opposed to vortex separation that happens when the system is conditioned not to couple to the stable modes. With no stable modes, the $x$-directed, initial magnetic fields are highly folded by the violent and energetic eddies.

This brute force numerical experiment provides a visual display that confirms the comprehensive and technical analysis presented earlier in this paper: vortex mergers, energetics, and cascades are all drastically different when the stable modes are not available to the system.

A further quantitative analysis of the standard and stable-mode-decoupled simulations is presented in Fig.~\ref{fig:f15}. In the latter simulation, the energy fluxes show enhancement in their levels by orders of magnitude, compared to the standard simulation, despite both evolving from identical flux spectra at the early stage.  The small-scale dissipation length-scale is also pushed to further smaller scales because of larger turbulent energy in the absence of stable modes.

\section{General implications}\label{sec:sec8}
Now we assess the detailed energetics shown in the preceding sections in relation to broader understanding and implications.

\subsection{Imprints of instability-scale flow at small scales}
A key aspect of the magnetic energy transfer is its nonlocality, in contrast to homogeneous isotropic turbulence, represented, for example, by the Kolmogorov spectrum.  Figures~\ref{fig:f4} and \ref{fig:f3}(b) have a pronounced diagonal feature indicative of a nonlocal interaction.  For each pair of red and blue cells across the diagonal, energy is exchanged via an interaction that is dominated by a single mode of the large-scale vortex flow arising from the Kelvin-Helmholtz instability.  Note that similar nonlinear interactions have been found in other turbulence where a system-size vortex is externally stirred.\cite{alexakis2005prl} The coupling in Fig.~\ref{fig:f4} of two small-scale magnetic modes with a large-scale flow is intrinsically nonlocal.  However, the energy exchanged tapers off significantly after a decade in wavenumber along the diagonal.  This indicates that an interaction that is intrinsically nonlocal is largely confined within a limited wavenumber range---a phenomenon enforced via energy-removal by the stable modes.

\subsection{Stable modes vs. Kolmogorov wavenumber and dissipation-range turbulence}

The Kolmogorov wavenumber $k_d$, where the turbulent energy cascade begins getting significantly attenuated because of small-scale energy dissipation, is directly related to the energy injection rate $\epsilon_f$ at large scales: $k_d = (\epsilon_f/d^3)^{1/4}$, where $d$ is a coefficient related to small-scale dissipation, such as viscosity or resistivity. As the energy cascade processes do occur in the shear-flow turbulence considered here, $k_d$ approximately delineates the dissipation range from the larger scales of turbulence. The energy injection rate is normally the energy $Q_1$ that is provided to the fluctuation spectrum by unstable modes. However, when stable modes are significantly excited via nonlinear processes, they act as large-scale turbulent sinks, and thus efficiently remove energy $Q_2$ from the fluctuation spectrum, steepening the mean-flow gradient.  Only the remaining energy $Q_1-Q_2 \approx \epsilon_f$ is then available to cascade nonlinearly to smaller scales. The energy injection rate for a simulation with $d=\nu=\eta=(1/500) a U_0$ can be determined from Fig.~\ref{fig:f12}(b) to be $\epsilon_f \approx 0.5 \times 10^{-3} U_0^3/a$. Taking contributions of stable modes $Q_2$ into account, the predicted Kolmogorov wavenumber is then $k_d \approx 15.8 a^{-1}$. This prediction is confirmed in Fig.~\ref{fig:f7}. Neglecting the energetics of stable modes $Q_2$ yields $k_d \approx 38.8$. The stable modes reduce the energy input rate $\epsilon_f \approx Q_1-Q_2$ to the small-scale energy cascade channel, and hence the cascade attenuates at a larger length scale.

The exponential fall-off of the MHD energy fluxes in spectral space in Fig.~\ref{fig:f7} also suggests that a simple model of the energy fluxes,\cite{pao1965} may be applicable. To test such a prospect, we follow the assumption that the spectral energy fluxes, in inertial and dissipative ranges, obey $\Pi(k) = p\, k^{-\beta}\exp{(-c k^\alpha)}$, where $c$ and $p$ are independent of the wavenumber $k$; this implies that the energy flux need not be constant, owing to energy absorption by dissipative physics, as opposed to what one would have in a strict inertial range ($\alpha=\beta=0$).\cite{terry2009diss, terry2012diss, hu2018diss}
Fitting such a profile to the energy fluxes in Fig.~\ref{fig:f7} (with a standard choice $\beta=0$ to allow a constant energy flux at larger scales), we find that the exponent $\alpha$ for the total magnetic energy flux at smaller scales $k_0 > 10$, $\Pi^{\mathrm{all}}_{\vecb>} = \Pi^{\vecu<}_{\vecb>} + \Pi^{\vecu>}_{\vecb>} + \Pi^{\vecb<}_{\vecb>}$, is very close to $4/3$, which matches with the widely recognized theoretical prediction of Ref.~\cite{pao1965}. The exponent $\alpha$ for the total kinetic energy flux $\Pi^{\mathrm{all}}_{\vecu>} = \Pi^{\vecu<}_{\vecu>} + \Pi^{\vecb<}_{\vecu>} + \Pi^{\vecb<}_{\vecu>}$, however, we find, is around $1/2$, which we are unable to explain with such a simple model; it is possible that the nonlinearly excited stable modes have some impact on this exponent.

The effect of stable modes on Kolmogorov dissipation length scale is substantiated also by Figs.~\ref{fig:f14} and \ref{fig:f15}. There, the stable-mode-decoupled simulation shows orders of magnitude of enhancement in the turbulent energy fluxes, which fall off only at much larger wavenumber than the fluxes do in the standard simulation.  This result is consequential.  Because the small-scale energy fluxes are very large when the stable modes are absent, traditional Kolmogorov-like scaling arguments of energy cascade are expected to fail, as, there, the energy injection rate to the nonlinear cascade is equated to the rate of energy withdrawl by the unstable modes from the mean shear-flow. Incorporating the energy reversal by the stable modes can make the scaling arguments succeed. \\

\subsection{Thermodynamic irreversibility and stable modes}
It is not unreasonable to assume that the wavenumber of the fastest-growing mode corresponds to the peak of the fluctuation spectrum. However, in gyrokinetic simulations of drift-wave turbulence, a shift of the spectrum peak to lower wavenumber is commonly observed, and is not fully understood.
In Figs.~\ref{fig:f2}(a) and \ref{fig:f9}, we find that the wavenumber $k_x=0.4$ where the perturbations linearly grow the fastest depletes energy from the fluctuation spectrum and deposits in the mean background gradients---a finding that is counter intuitive, at first. Further examination in Fig.~\ref{fig:f9} reveals that the stable-mode amplitude exceeds the unstable-mode amplitude at $k_x=0.4$. This does not violate the laws of thermodynamics, because the energy transfer summed over all wavenumbers is directed from the mean gradient to the fluctuation spectrum, in accordance with the breakdown of time-reversal symmetry due to visco-resistive effects.  This forward flow of energy from the mean gradient to the fluctuation spectrum is evident at $k_x=0.2$, where the spectrum peaks.

\section{Conclusions}\label{sec:sec9}

This paper has systematically examined energy transfer processes in MHD turbulence driven by shear-flow instability, quantifying the effect of linearly stable roots of the dispersion relation \cite{chandrashekhar1961}, which are nonlinearly excited to a significant level, thus modifying the landscape of fluctuation source and sink at large scales.  
The role of nonlinearly excited stable modes as a fluctuation energy sink has been extensively examined for fusion-relevant microinstability, where the prompt loss of energy at the largest scales allows the formulation of saturation balances that exclude extended scale ranges of cascaded energy \cite{terry2018, li2022}.  In this paper, we have examined the effect of stable modes on the energy cascades, quantifying energy transfer rates under an eigenmode decomposition that tracks energy transfer between finite-amplitude-induced eigenmodes.
The canonical direct cascade of kinetic and magnetic energies \cite{biskamp2003} remains in force despite the presence of stable eigenmodes. However, the energy carried in the cascade is dramatically reduced compared to the energy fed into
the fluctuation spectrum by the unstable modes, with most of that unstable-mode energy returned to the mean flow by the stable modes. This fraction of energy return ranges from around $75\%$ to $97\%$, depending on the strength of the imposed magnetic field and on the magnetic Prandtl number. 
The cascading energy flux to small scales does dissipate at small visco-resistive scales, although such energy flux exponentially attenuates at small scales, dictated by the amplitudes of the stable modes at large scales, as such modes directly control the energy input rate into the small-scale cascade.

The energy nonlinearly transferred from the unstable modes and deposited almost in its entirety into the stable modes has two distinct and dominant channels. These arise from the nonlinear interactions, first, among the fluctuations composed of the discrete modes $T_{j\mathrm{dd}}$ and, second, the nonlinear interactions between the fluctuations of the discrete and the continuum modes $T_{j\mathrm{dc}}$. The discrete modes are the unstable $j=1$ and conjugate-stable $j=2$ eigenmodes of the ideal linear operator of the Kelvin-Helmholtz instability. The continuum modes are marginally stable and occur across a continuum of frequencies and across all scales, even beyond the instability scale. The strength of the interaction among the fluctuations of the continuum modes $T_{j\mathrm{cc}}$, which in principle affects the evolution of the unstable and stable modes, is found to be negligibly small, ranging from $2\%$ to $23\%$ of $T_{j\mathrm{dd}}$ or $T_{j\mathrm{dc}}$. The interaction term $T_{j\mathrm{dd}}$ is found to always take energy away from the unstable modes and channel it away almost in its entirety to the stable modes.  On the other hand, the nonlinear transfer $T_{j\mathrm{dc}}$ always takes energy from the stable modes and provides it nonlinearly to the unstable modes. Nevertheless, $T_{j\mathrm{dd}}$ is always larger than $T_{j\mathrm{dc}}$ in magnitude. Thus the nonlinearity transfers a net positive energy from unstable to stable modes.

Based upon the strength of the nonlinear interactions between the fluctuations classified in the eigenmode basis, a reduced-order model for the subgrid-scale turbulence generated by the Kelvin-Helmholtz instability may be possible. A self-consistent subgrid-scale model could involve ignoring the eddy-eddy nonlinearity beyond a prescribed cutoff wavenumber.\cite{marston2016} This cutoff wavenumber could be chosen as low as the inverse of the shear-width of the mean profile, which is where the instability ceases to exist (i.e., $|k_x| \sim 1$).

For the first time, we have built and tested a general quasilinear model of Kelvin-Helmholtz-instability-driven turbulence, motivated by prevailing hypothesis of instability saturation\cite{fuller2019, pessah2006, goodman1994, garaud2018, barker2019}. We have discovered that even a model where all scales of turbulence are solved exactly, including all eigenmodes---both unstable and continuum modes---but removing just their coupling to the stable modes, fails to reproduce even the primary features of the Kelvin-Helmholtz turbulence. For instance, the usual large-scale vortex merger events in $2$D are missed, and instead an explosive separation of large-scale vortices is seen, in addition to a rapid spreading of turbulence away from the shear layer. Such a dramatic difference in the structures and levels of turbulence, along with enhanced turbulent energy fluxes, when the couplings to the stable modes are analytically removed, confirm that the stable modes act as a large-scale energy sink, and thus tame the turbulence near the narrow region of the shear layer. This finding has consequences in modeling efforts of shear-flow turbulence in fusion plasmas when the zonal flows and streamers go unstable.

It has not escaped our understanding that, given the critical role of stable modes with respect to the structures and energetics of the Kelvin-Helmholtz turbulence, studies of MHD turbulence in reconnection-driven sheared outflows\cite{schekochihin2022} may benefit from the investigation of stable modes in such settings, and simpler models of scaling and cascade rates may be informed from such analyses. Although the stable modes considered in this paper are of the shear-flow instability, other instabilites too have stable modes, for instance, the tearing instability,\cite{hu2018diss} which can co-exist with the Kelvin-Helmholtz instability. The stable modes, as here in shear-flow turbulence, may deplete the fluctuation energy there as well, thus potentially lowering the small-scale energy cascade rate and affecting the spectral index of the fluctuation power spectrum. More work needs to be carried out in the future to assess the impact of stable modes on, for example, MHD energy fluxes and the breaking of energy cascade.\cite{dong2022}

\begin{acknowledgments}
This material is based upon work funded by the Department of Energy [DE-SC0022257] through the NSF/DOE Partnership in Basic Plasma Science and Engineering. We are grateful to K.~Burns and the Dedalus developers for technical help. We thank D.~Mitra for suggesting the diagrammatic data presentation. Useful discussions with Y.M.~Huang are acknowledged. A.E.F. acknowledges support from NASA HTMS grant 80NSSC20K1280, and from the George Ellery Hale Postdoctoral Fellowship in Solar, Stellar and Space Physics at the University of Colorado, Boulder. E.H.A. is supported by a CIERA Postdoctoral Fellowship. The simulations were performed using the XSEDE/ACCESS supercomputing resources via Allocation No.~TG-PHY130027. 

The data that support the findings of this study are available from the corresponding author upon reasonable request.
\end{acknowledgments}

\setcounter{equation}{0}
\setcounter{figure}{0}
\setcounter{table}{0}
\makeatletter
\renewcommand{\theequation}{A\arabic{equation}}
\renewcommand{\thefigure}{A\arabic{figure}}

\section*{Appendix: Anti-symmetric $S$-transfer-function}

The anti-symmetry property of the wavenumber-to-wavenumber ($S$-) transfer function
\begin{equation}
    S^{\vecc}_{\veca}(k_x\vert k_x'')  = - S^{\veca}_{\vecc}(k_x''\vert k_x),
\end{equation}
 will be proved here.

To begin, take Eq.~\eqref{eq:scakxkxpp}
\begin{equation} \label{eq:scakxkxppappendix}
    S^{\vecc}_{\veca}(k_x\vert k_x'') =  \real \Bigg\{\Big\langle \veca(-k_x) \cdot \Big[\vecb(k_x') \cdot \nabla'' \vecc(k_x'') \Big] \Big\rangle_z \Bigg\}.
\end{equation}
where the triadic interaction involves $\veca, \vecb,$ and $\vecc$ at wavenumbers $k_x, k_x', $ and $k_x''$ such that $-k_x+k_x'+k_x''=0$. Following the analogy with Eq.~\eqref{eq:scakxkxppappendix}, we compose the expression for $S^{\veca}_{\vecc}(k_x''\vert k_x)$, with the constraint $-k_x+k_x'+k_x''=0$ still applied, as
\begin{equation}
\begin{aligned}
    S^{\veca}_{\vecc}(k_x''\vert k_x) &=  \real \Bigg\{\Big\langle \vecc(-k_x'') \cdot \Big[\vecb(-k_x') \cdot \nabla \veca(k_x) \Big] \Big\rangle_z \Bigg\} \\
    &= \real \Bigg\{\Big\langle \vecc(k_x'') \cdot \Big[\vecb(k_x') \cdot \nabla \veca(-k_x) \Big] \Big\rangle_z \Bigg\} \\
    &=  S^{\veca}_{\vecc}(-k_x''\vert -k_x),
\end{aligned}
\end{equation}
which simply means that the energy transfer from $k_x$ to $k_x''$ is same as the energy transfer from $-k_x$ to $-k_x''$, i.e., conjugate symmetry, as was shown also for the net energy transfer in Eq.~\eqref{eq:conservationT1}.

With the expressions for $S^{\vecc}_{\veca}(k_x\vert k_x'')$ and $S^{\veca}_{\vecc}(k_x''\vert k_x)$ at hand, we now show that they hold an antisymmetry property: $S^{\vecc}_{\veca}(k_x\vert k_x'') = -S^{\veca}_{\vecc}(k_x''\vert k_x)$. To prove such, let us evaluate $S^{\vecc}_{\veca}(k_x\vert k_x'') + S^{\veca}_{\vecc}(k_x''\vert k_x)$ below
\begin{equation} \label{eq:append1}
    S^{\vecc}_{\veca}(k_x\vert k_x'') + S^{\veca}_{\vecc}(k_x''\vert k_x) = \real \Big\langle \veca^\ast(k_x) \cdot \Big[\vecb(k_x') \cdot \nabla \vecc(k_x'') \Big] \Big\rangle_z  + \real \Big\langle \vecc^\ast(k_x'') \cdot \Big[\vecb^\ast(k_x') \cdot \nabla \veca(k_x) \Big] \Big\rangle_z
\end{equation}

The first term on the right-hand side of Eq.~\eqref{eq:append1} is
\begin{equation} \label{eq:a4a}
\real \Big\langle\veca^\ast(k_x) \cdot \Big[\vecb(k_x') \cdot \nabla \vecc(k_x'') \Big] \Big\rangle_z
=\real \Big\langle \veca^\ast \cdot (B_x'  i k_x'' + B_z' \partial_z) \vecc''  \Big\rangle_z,
\end{equation}
and the second term on the right-hand side of Eq.~\eqref{eq:append1} is
\begin{equation}\label{eq:a4c}
\begin{aligned}[b]
\real \Big\langle\vecc^\ast(k_x'') \cdot \Big[\vecb^\ast(k_x') \cdot \nabla \veca(k_x) \Big]\Big\rangle_z
&=\real \Big\langle\vecc''^\ast \cdot (B_x'^\ast  i k_x + B_z'^\ast \partial_z) \veca\Big\rangle_z\\
&=\real \Big\langle\vecc'' \cdot (-B_x'  i k_x + B_z'\partial_z) \veca^\ast \Big\rangle_z.
\end{aligned}
\end{equation}

Substituting the expressions from Eqs.~\eqref{eq:a4a} and \eqref{eq:a4c} in Eq.~\eqref{eq:append1},
\begin{equation} \label{eq:appendixlastzero}
\begin{aligned}[b]
    S^{\vecc}_{\veca}(k_x\vert k_x'') + S^{\veca}_{\vecc}(k_x''\vert k_x) 
    &= \real \Big\langle \veca^\ast \cdot (B_x'  i k_x'' + B_z' \partial_z) \vecc''  \Big\rangle_z + \real \Big\langle\vecc'' \cdot (-B_x'  i k_x + B_z'\partial_z) \veca^\ast \Big\rangle_z\\
    &= \real \Big\langle  B_x'  i (k_x'' - k_x) \veca^\ast \cdot  \vecc''  \Big\rangle_z 
    + \real \Big\langle B_z' \veca^\ast \cdot  \partial_z \vecc''  + B_z' \vecc'' \cdot \partial_z \veca^\ast \Big\rangle_z\\
    &= -\real \Big\langle  ik_x' B_x'  \veca^\ast \cdot  \vecc''  \Big\rangle_z 
    + \real \Big\langle B_z' \partial_z (\veca^\ast \cdot  \vecc'' ) \Big\rangle_z\\
    &= -\real \Big\langle  ik_x' B_x'  \veca^\ast \cdot  \vecc''  \Big\rangle_z 
    - \real \Big\langle (\partial_z B_z') \veca^\ast \cdot  \vecc''  \Big\rangle_z\\
    &= -\real \Big\langle  (ik_x' B_x' + \partial_z B_z')  \veca^\ast \cdot  \vecc''  \Big\rangle_z \\
    &= -\real \Big\langle  [\nabla' \cdot \vecb(k_x')] \veca^\ast \cdot  \vecc''  \Big\rangle_z \\
    &= 0.    
\end{aligned}
\end{equation}
In the second last line of Eq.~\eqref{eq:appendixlastzero}, it can be seen that divergence of the vector field $B$ appears. Since all the vectors fields---velocity and magnetic fields---are divergenceless in this study, we obtain null at the end. 
Thus the anti-symmetry property of the wavenumber-to-wavenumber ($S$-) transfer function
\begin{equation}
    S^{\vecc}_{\veca}(k_x\vert k_x'')  = - S^{\veca}_{\vecc}(k_x''\vert k_x),
\end{equation}
is proved. It can be a fruitful exercise for the reader to repeat this proof in a fully periodic system,\cite{verma2019} where the proof requires only a couple of lines of equation.

\end{document}